\documentclass[fleqn,usenatbib]{mnras}

\usepackage[T1]{fontenc}
\usepackage{xcolor}
\usepackage{cuted}
\usepackage{booktabs}

\DeclareRobustCommand{\VAN}[3]{#2}
\let\VANthebibliography\thebibliography
\def\thebibliography{\DeclareRobustCommand{\VAN}[3]{##3}\VANthebibliography}

\usepackage{graphicx}	
\usepackage{amsmath}	
\usepackage{amssymb}	
\usepackage{subcaption}
\usepackage{physics}

\usepackage[nolist]{acronym}
\newacro{CMB}{cosmic microwave background}
\usepackage{newtxtext,newtxmath}

\newcommand{\rmhh}{{\mathrm{hh}}}
\newcommand{\rmgg}{{\mathrm{gg}}}
\newcommand{\rmg}{{\mathrm{g}}}
\newcommand{\mathrms}{{\mathrm{s}}}
\newcommand{\mathrmc}{{\mathrm{c}}}

\newcommand{\mathrmR}{{\mathrm{R}}}

\title[Dark Quest I.1]{Galaxy clustering from the bottom up: A Streaming Model emulator I}

\author[C. Cuesta-Lazaro et al.]{
Carolina Cuesta-Lazaro$^{1,2}$,\thanks{E-mail: carolina.cuesta-lazaro@durham.ac.uk}
Takahiro Nishimichi$^{3,4}$,
Yosuke Kobayashi$^{5}$,
Cheng-Zong Ruan$^{1}$,\
\newauthor 
Alexander Eggemeier$^{8}$ \thanks{Argelander Fellow},
Hironao Miyatake$^{6,4}$,
Masahiro Takada$^{4}$,
Naoki Yoshida$^{7,4}$,
Pauline Zarrouk$^{9}$,\
\newauthor
Carlton M. Baugh,$^{1,2}$, 
Sownak Bose$^{1}$ and Baojiu Li$^{1}$\\
$^{1}$Institute for Computational Cosmology, Department of Physics, Durham University, South Road, Durham DH1 3LE, UK\\
$^{2}$Institute for Data Science, Durham University, South Road, Durham DH1 3LE, UK\\
$^{3}$ Center for Gravitational Physics, Yukawa Institute for Theoretical Physics, Kyoto University, Kyoto 606-8502, Japan\\
$^{4}$ Kavli Institute for the Physics and Mathematics of the Universe
(WPI),\\ 
The University of Tokyo Institutes for Advanced Study (UTIAS),
The University of Tokyo, Chiba 277-8583, Japan\\
$^{5}$ Department of Astronomy/Steward Observatory, University of Arizona, 933 North Cherry Avenue, Tucson, AZ 85721-0065, USA\\
$^{6}$ Kobayashi-Maskawa Institute for the Origin of Particles and the Universe (KMI), Nagoya University, Nagoya, 464-8602, Japan\\
$^{7}$ Department of Physics, The University of Tokyo, 7-3-1 Hongo, Bunkyo, Tokyo 113-0033, Japan\\
$^{8}$ Argelander Institut f\"ur Astronomie der Universit\"at Bonn, Auf dem H\"ugel 71, 53121 Bonn, Germany \\
$^{9}$ Sorbonne Universit\'e, Universit\'e Paris Diderot, Sorbonne Paris Cit\'e, CNRS,
Laboratoire de Physique Nucléaire et de Hautes Energies (LPNHE), \\ 4 place Jussieu, F-75252, Paris
Cedex 5, France}

\date{Accepted XXX. Received YYY; in original form ZZZ}

\pubyear{2022}

\begin{document}
\label{firstpage}
\pagerange{\pageref{firstpage}--\pageref{lastpage}}
\maketitle

\begin{abstract}
In this series of papers, we present a simulation-based model for the non-linear clustering of galaxies based on separate modelling of clustering in real space and velocity statistics. In the first paper, we present an emulator for the real-space correlation function of galaxies, whereas the emulator of the real-to-redshift space mapping based on velocity statistics is presented in the second paper. Here, we show that a neural network emulator for real-space galaxy clustering trained on data extracted from the \textsc{Dark Quest} suite of N-body simulations achieves sub-per cent accuracies on scales $1 < r < 30 $ $h^{-1} \,\mathrm{Mpc}$, and better than $3\%$ on scales $r < 1$ $h^{-1}\mathrm{Mpc}$ in predicting the clustering of dark-matter haloes with number density $10^{-3.5}$ $(h^{-1}\mathrm{Mpc})^{-3}$, close to that of SDSS LOWZ-like galaxies. The halo emulator can be combined with a galaxy-halo connection model to predict the galaxy correlation function through the halo model. We demonstrate that we accurately recover the cosmological and galaxy-halo connection parameters when galaxy clustering depends only on the mass of the galaxies' host halos. Furthermore, the constraining power in $\sigma_8$ increases by about a factor of $2$ when including scales smaller than $5$ $h^{-1} \,\mathrm{Mpc}$. However, when mass is not the only property responsible for galaxy clustering, as observed in hydrodynamical or semi-analytic models of galaxy formation, our emulator gives biased constraints on $\sigma_8$. This bias disappears when small scales ($r < 10$ $h^{-1}\mathrm{Mpc}$) are excluded from the analysis. This shows that a vanilla halo model could introduce biases into the analysis of future datasets.
\end{abstract}




\section{Introduction}
The large scale structure (LSS) of the Universe as shown by three-dimensional galaxy maps carries a wealth of information which can be used to constrain theories of gravity. In particular, we can use the clustering properties of the LSS to address some of the most pressing questions faced by the standard cosmological model, such as what drives the accelerated expansion of the Universe and what is the dark matter. Ongoing and future surveys, such as the Dark Energy Spectroscopic Instrument (DESI)  \citep{2016arXiv161100036D}, the Subaru Prime Focus Spectrograph (PFS)  \citep{SubaruPFS}, and the space-based mission Euclid \citep{2011arXiv1110.3193L} will provide LSS maps of unprecedented statistical precision. The challenge for cosmologists now is to develop statistical methods that are accurate enough to match the precision of the data, so that we can extract all of the valuable information on gravity and cosmology contained in the LSS.

Galaxy clustering provides us with a means to  constrain the  cosmological model  through  late-Universe measurements. This enables us to carry out a consistency check by comparing cosmological constraints derived from observations of the early and late Universe and determining whether or not the results are consistent with the evolution expected in a $\Lambda$-cold dark matter ($\Lambda$CDM)  model. Inconsistencies of more than $2\text{-}3 \,\sigma$ have been found when comparing the matter clustering strength, $\sigma_8$, inferred from the early Universe through cosmic microwave background (CMB) measurements, with the estimate from the late Universe, as deduced from both weak gravitational lensing and galaxy clustering \citep{10.1093/mnras/stw2665,2019PASJ...71...43H, PhysRevD.105.023520, PhysRevD.105.043517}. Late Universe probes prefer a smaller value of $\sigma_{8}$ and hence a lower degree of structure formation than is expected from \ac{CMB} observations (see \citealt{osti_1862246} for a detailed discussion of the so-called $\sigma_8-S_8$ tension). Reducing the uncertainties on the estimated cosmological parameters would help to determine if the observed tension is the result of systematics, statistical bad luck, or even the imprint of new physics that is yet to be discovered. 

Given a 3-D galaxy field, one could aim to infer the cosmological parameters directly at the field level by comparing the gridded number density of the observed galaxies with that expected from a model \citep{10.1093/mnrasl/slab081, Elsner_2020}. There are two different sets of variables that play a role in determining the expected number density. On the one hand, the random phases of the initial conditions determine where the initial seeds that gave rise to the observed large scale structure were located. On the other hand, the cosmological parameters influence how these seeds will collapse through gravitational evolution. However, jointly constraining the initial random phases and the cosmological parameters is a very challenging task. To avoid this difficulty, we define summary statistics of the 3-D galaxy maps that aim to reduce the stochasticity of sampling the initial conditions, whilst preserving as much information as possible about the cosmological parameters. 

If the galaxy field were a Gaussian random field, its two-point statistics (the power spectrum or the two-point correlation function) would contain all information there is in the full 3-D maps. But while the density field at high redshift is indeed close to Gaussian over a wide range of scales, nonlinear gravitational evolution produces non-Gaussianity. Given that the mass overdensity $ \left({\rho(x)-\bar{\rho}(x)}\right)/\bar{\rho}(x)$, where $\rho$ is the mass density, is bounded at low values by $-1$, since a region of the Universe cannot have a negative density, the distribution of $\delta$ values must develop skewness as the density contrast grows. Finding alternative summary statistics to supplement the constraints obtained from the two-point functions is currently an active area of research (see, for instance, studies on the bispectrum, \citealt{Hahn_2020}, the scattering transform, \citealt{2022PhRvD.105j3534V}, and density split statistics for galaxy clustering \citealt{10.1093/mnras/stab1654}).

An alternative way to maximise the information that is extracted from cosmological surveys is by modelling the cosmological dependence of small scale clustering. Although the statistical precision of data on small scales is higher than that on large scales, most studies that rely on perturbation theory \citep[e.g.][]{2021JCAP...03..100C} to model the dependence of two-point functions on cosmology restrict their analysis to pair separations larger than $\approx 30$ $h^{-1}\mathrm{Mpc}$. On smaller scales, the accuracy of perturbation theory breaks down rapidly, and its use introduces biases in the inferred cosmological parameters. The additional constraining power of small scales was demonstrated by \citet{Zhai_2019} who showed how the constraints on the growth rate of structure, $f$, and the clustering amplitude, $\sigma_8$, increase monotonically as smaller scales are added to the analyses.

To obtain fully non-linear predictions for the properties of the large-scale structure and recover all the cosmological information contained in the small-scale clustering, we must resort to N-body simulations \citep{Kuhlen:2012ft}. N-body simulations have been widely used as cosmic laboratories to test the precision and robustness of analytical methods for the large-scale structure (e.g., \citealt{2009PhRvD..80d3531C, PhysRevD.91.023508, 10.1093/mnras/staa2249}), together with the effects of systematic errors in our measurements. Over the past decade, advances in computing have allowed us to produce a large enough number of dark matter only N-body simulations covering a significant fraction of the cosmological parameter space, which allows us to use the simulations themselves as predictive models that directly constrain the cosmological parameters. These simulations both need to be large enough to reduce sample variance, and have a high enough resolution to resolve the tracers that will be surveyed.

Moreover, in order to compare the outcomes of dark matter only simulations to the
observed distribution of galaxies we have to model the connection between dark matter halos
and galaxies (see \citealt{2018ARA&A..56..435W} for a review on this topic). Uncertainties in the galaxy-halo connection can limit the amount of information that we can extract from small scale clustering. We would like to use flexible models that can reproduce clustering in different scenarios of galaxy formation, whilst still being able to recover cosmological information after marginalising over the free parameters of the galaxy-halo connection model. Here, we use the empirical model of the halo occupation distribution (HOD) \citep{Benson_2000,2005ApJ...633..791Z}, based on estimating the probability that a given halo hosts a galaxy.

Over the past few years, several studies \citep{Zhai_2019,2019MNRAS.490.1870L,PhysRevD.102.063504,2021arXiv211102419M} have shown how N-body simulations can be leveraged to extract small scale information. Solving the inverse problem, estimating the posterior over the cosmological parameters given the observed clustering, would require the order of $\mathcal{O}(10^6)$ N-body simulations to perform Bayesian inference with Markov Chain Monte Carlo. Therefore, most studies rely on modelling the dependence of the two-point correlation function on cosmology with surrogate models that are trained on a small set of $\mathcal{O}(100)$ N-body simulations \citep{Zhai_2019,2019MNRAS.490.1870L, PhysRevD.102.063504}. The surrogate models are orders of magnitude faster than the original N-body simulations and can then be used to sample the posterior of cosmological parameters. 

For instance, \citet{PhysRevD.102.063504} developed an N-body version of the halo model for the galaxy power spectrum by training a neural network to reproduce the dark matter halo clustering properties in Fourier space.  \citet{Zhai_2019} and \citet{2022arXiv220311963Y} followed a different route by emulating galaxy clustering as both a function of cosmology and galaxy-halo connection parameters with Gaussian processes \citep{10.5555/1162254}. Alternatively, \citet{2019MNRAS.490.1870L} developed the so-called cosmological evidence modelling (CEM) method. \citet{2019MNRAS.490.1870L}  used N-body simulations to compute the evidence of the data as a function of cosmology after marginalising over the HOD parameters, which can then be used to sample the posterior distribution over the cosmological parameters. In this way, the authors do not have to account for the errors introduced by the surrogate model although errors in the emulation of the likelihood function would still impact the inference. However, this approach does not yield joint constraints on the galaxy-halo connection and cosmological parameters, since the HOD parameters are marginalised over. 

These simulation-based methods currently produce the tightest constraints on the combination $f\sigma_8$ \citep{Lange_2021, Kobayashi:2021oud, 2022arXiv220311963Y, 2022arXiv220308999Z} when confronted with observations. Interestingly, all studies find values for the combination $f\sigma_8$ that are lower than those obtained from the CMB. The current challenge for emulator-based approaches is to both make sure that theoretical predictions are on a par with the statistical errors expected from future surveys, and that the modelling of how galaxies populate dark matter halos does not introduce biases into the analysis from small-scale clustering.

In this series of papers we build emulators for both real space clustering and pairwise velocity statistics \citep{Peebles1980, Fisher1995}; the latter determine the mapping between real and redshift space clustering. In this way, we will be able to combine constraints from clustering measurements and estimates of peculiar velocities, obtained through either the kinetic Sunyaev-Zeldovich effect \citep{10.1093/mnras/190.3.413} (see \citet{PhysRevD.104.043502} for a recent measurement) or through peculiar velocity surveys \citep{10.1093/mnras/stz901}, to obtain more precise constraints on the cosmological parameters. Peculiar velocity surveys and redshift space distortions have been shown to be specially complementary to test gravity theories \citep{Kim_2020}.

In this first paper of the series we focus on modelling small scale galaxy clustering in real space, improving on the emulators presented in \citet{Nishimichi:2018etk} in terms of both accuracy and speed. We show how a combination of neural networks trained using the predictions of N-body simulations and the halo model can produce extremely accurate predictions for the clustering of galaxies over a wide range of pair separations, $0.01 < r < 150$ $h^{-1}\,\mathrm{Mpc}$, as opposed to the range $r < 30$ $h^{-1}\,\mathrm{Mpc}$, covered by previous emulators in configuration space \citep{Zhai_2019,PhysRevD.102.063504}. This allows us to compute the likelihood using the full shape of the two-point correlation function, spanning the behaviour of the one- and two-halo terms. Finally, we demonstrate the limitations of the current implementation of the halo model to recover unbiased constraints when an assembly bias signal \citep{2018ARA&A..56..435W} is present in the data to be analysed.

This paper is organised as follows. In Section~\ref{sec:background}, we introduce the theoretical model of redshift-space clustering. In Section~\ref{sec:sim_suite} we describe the N-body simulations used to train the emulator. In Section~\ref{sec:halo_to_galaxy} we describe the halo model approach for predicting galaxy clustering. In Section~\ref{sec:neural_nets} we present neural network emulators trained to reproduce the clustering of dark matter halos and their abundance, and show how they can be combined with the halo model to accurately reproduce the clustering of galaxies. Section~\ref{sec:inverse_problem} focusses on solving the inverse problem to obtain unbiased posterior distributions over the cosmological parameters. In particular, we show the limitations of the halo model in recovering unbiased constraints on $\sigma_8$ when assembly bias is present and scales smaller than $10  h^{-1}\, \mathrm{Mpc}$ are included in the likelihood. Finally, we present our conclusions in Section \ref{sec:discussion}.

\section{Theoretical Background}
\label{sec:background}

The two-point correlation function, $\xi^R (r)$, quantifies clustering as the excess probability of finding a pair of galaxies at a given separation, compared with a random distribution of galaxies. The two-point correlation function is defined as
\begin{equation}
\xi^R (r) = \langle \delta(\mathbf{x}) \delta(\mathbf{x} + \mathbf{r}) \rangle,
\end{equation}
where $\delta = \left({\rho(x)-\bar{\rho}(x)}\right)/\bar{\rho}(x)$ is the density contrast and $\bar{\rho}$ is the mean density. When assuming statistical isotropy and homogeneity, $\xi^R$ depends only on pair separation, $r$. 

In redshift surveys, we measure the angular positions of galaxies in the sky and their redshift. Then, the angular coordinates can be converted to comoving distances by assuming a cosmology through the angular diameter distance. If we assume that galaxies are at rest, as the photons emitted by galaxies travel towards us through an expanding universe, their wavelengths stretch accordingly, producing the redshift effect. 
We can translate this redshift into a comoving distance by introducing the Hubble factor, $H(z)$:
\begin{equation}
\label{eq:dist2red}
    r(z) = \int_0^z \frac{d z^\prime}{H(z^\prime)},
\end{equation}
where $r(z)$ is the comoving distance to the galaxy, and we have used the natural unit where the speed of light $c=1$.

Nevertheless, there are several effects related to the distorted way in which we observe the Universe that complicate this simple picture. In fact, much of the information used to constrain cosmology from 3-D galaxy maps does not come directly from the underlying comoving map of galaxy positions, but from the distortion effects that alter this map. 

In particular, galaxies move because of the gravitational pull generated by the inhomogeneous distribution of matter around them. If a source that emits light moves, the wavelength of its light becomes further redshifted because of the Doppler effect. If we ignored this effect, then we would infer the wrong position, $\mathbf{s}$, given by
\begin{equation}
\mathbf{s} = \mathbf{r} + \frac{\mathbf{v}(\mathbf{r})\hat{z}}{\mathcal{H}}\hat{z},
\label{eq:rspositions}
\end{equation}
instead of the real position of the galaxy, $\mathbf{r}$ ,where $\mathbf{v}(\mathbf{r})$ is the peculiar velocity of the galaxy, $\mathcal{H} = a H(a)$ the comoving Hubble factor, where $a$ is the expansion factor, and the inferred distance, $\mathbf{s}$, is the redshift space position of the galaxy. 
Note that we have assumed that the observer is far away from the sources and therefore the line-of-sight direction can be fixed to a particular direction, regardless of the angular coordinates of the galaxy, which we arbitrarily set as the $\hat{z}$ axis. 

Due to peculiar motions of galaxies, we observe redshift space positions, $\mathbf{s}$, instead of the real space positions, $\mathbf{r}$, and thus we can only measure 
\begin{equation}
\xi^S (s_\perp, s_\parallel) = \langle \delta(\mathbf{x}) \delta(\mathbf{x} + \mathbf{s}) \rangle,
\label{eq:redshift_tpcf}
\end{equation}
which depends on both the pair separation, $s$, and its inclination with respect to the line of sight direction. Throughout, we denote the separations perpendicular and parallel to the line of sight by $s_\perp$ and $s_\parallel$, respectively.

The two-point correlation function of galaxies in redshift space has been used to obtain tight constraints on the cosmological parameters in a $\Lambda$CDM universe (e.g. \citealt{2013AJ....145...10D,2016AJ....151...44D}). The so-called redshift space correlation function, $\xi^S (s_\perp, s_\parallel) $, is a combination of real space clustering, $\xi^R(r)$,  and the probability of finding a pair of galaxies with a given relative velocity along the line of sight, also denoted as the pairwise velocity distribution, $P(v_\parallel|r_\parallel, r_\perp)$. This is summarised in the following equation for the streaming model of redshift space distortions \citep{Fisher1995, PhysRevD.70.083007}
\begin{equation}
    \xi^S (s_\perp, s_\parallel) = \int d r_\parallel \left(1 + \xi^R(r) \right) P(s_\parallel - r_\parallel|r_\parallel, s_\perp).
    \label{eq:streaming}
\end{equation}

In \citet{10.1093/mnras/staa2249}, we show how the mapping between the real and redshift space can be accurately described by an analytical expression for $P(v_\parallel|r_\parallel, r_\perp)$, where $v_\parallel$ is the line of sight velocity in units of conformal $H(a)$. In this series of papers, we will model the cosmological dependence of the different ingredients of the streaming model: i) the two-point correlation function in real space, shown in this paper, and ii) the lowest four-order moments of the velocity field, needed to perform the real-to-redshift space mapping, as shown in \citet{10.1093/mnras/staa2249}. 

\section{The \textsc{Dark Quest} simulation suite}
\label{sec:sim_suite}
Here, we briefly describe \textsc{Dark Quest}, a suite of cosmological N-body simulations used to build emulators. 
A detailed description can be found in \citet{Nishimichi:2018etk}.

\begin{table*}
\centering
\caption{Comparison of the characteristics of the \textsc{DarkQuest} suite of simulations and those used to train clustering emulators in the literature. The mass of dark matter particles $M_{\mathrm{part}}$ has units of $\displaystyle (\Omega_{\rm m}/0.3) h^{-1}M_{\odot}$}
\label{tab:table_sims}
\begin{tabular}{ c c c c c c c}
 \hline
 \hline
 Simulation Suite & Code & $L_{\mathrm{box}}\,[h^{-1}\mathrm{Gpc}]$ & $N_\mathrm{part}$ & $M_\mathrm{part}$ & Halo Finder &  Reference   \\ 
  \hline
 \hline
 DarkQuest HR & \textsc{GADGET2}  & $1$ & $2048^3$ & $1.02 \times 10^{10}$ & \textsc{Rockstar} &  \citet{Nishimichi:2018etk} \\  
 DarkQuest LR & \textsc{GADGET2}   & $2$ & $2048^3$ & $8.158 \times 10^{10}$ & \textsc{Rockstar} & \citet{Nishimichi:2018etk} \\   
 AbacusSummit Base & \textsc{ABACUS}  & $2$ & $6912^3$ & $2.1 \times 10^9$  & CompaSO & \citet{10.1093/mnras/stab2484} \\
  Aemulus & \textsc{GADGET2}  & $1.05$ & $1400^3$ & $3.51 \times 10^{10}$ & \textsc{Rockstar} & \citet{DeRose_2019} \\
  \hline
\end{tabular}
\end{table*}

\subsection{N-body simulations}
\label{subsec:n_body_sim}
The \textsc{Dark Quest}  simulations were performed with $2048^3$ dark matter particles in $1\,h^{-1}\,\mathrm{Gpc}$ (hereafter high-resolution runs, denoted HR) or $2\,h^{-1}\,\mathrm{Gpc}$ (low-resolution runs, labelled LR) side-length boxes, using the \textsc{Gadget2}  N-body solver \citep{Gadget2:2005MNRAS.364.1105S}.
The mass resolutions of the HR and LR runs are $1.02 \times 10^{10}$ and $8.16 \times 10^{10} (\Omega_{\rm m}/0.3)\,h^{-1}\,M_{\odot}$, respectively. In Table \ref{tab:table_sims}, we show a comparison of the specifications of \textsc{Dark Quest} with those of other simulation suites that have been used to train clustering emulators in the literature \citep{Zhai_2019,2019MNRAS.490.1870L,PhysRevD.102.063504,2021arXiv211102419M}. \textsc{Dark Quest}, used int his work, has a higher resolution and a larger box size than Aemulus, but a lower resolution than AbacusSummit. In the future, it will be important to demonstrate the impact of differences in N-body codes (e.g.  \citealt{2021arXiv211209138G}), halo finders (e.g. \citealt{2022MNRAS.510.5500G}), and resolution on the cosmological parameters inferred using simulation-based methods.

The initial conditions were generated using second-order Lagrangian perturbation theory (2LPT, \citet{2LPTIC:2006MNRAS.373..369C}) and the redshift at which to generate the initial conditions was chosen depending on the cosmology and resolution  \citep{Nishimichi:2018etk}, with $z_{\mathrm{init}} \approx 59$ and $29$ adopted for the fiducial HR and LR simulations respectively.  Each simulation used different random number seeds to generate the initial conditions.

\begin{figure*}
    \centering
    \includegraphics[width=0.7\textwidth]{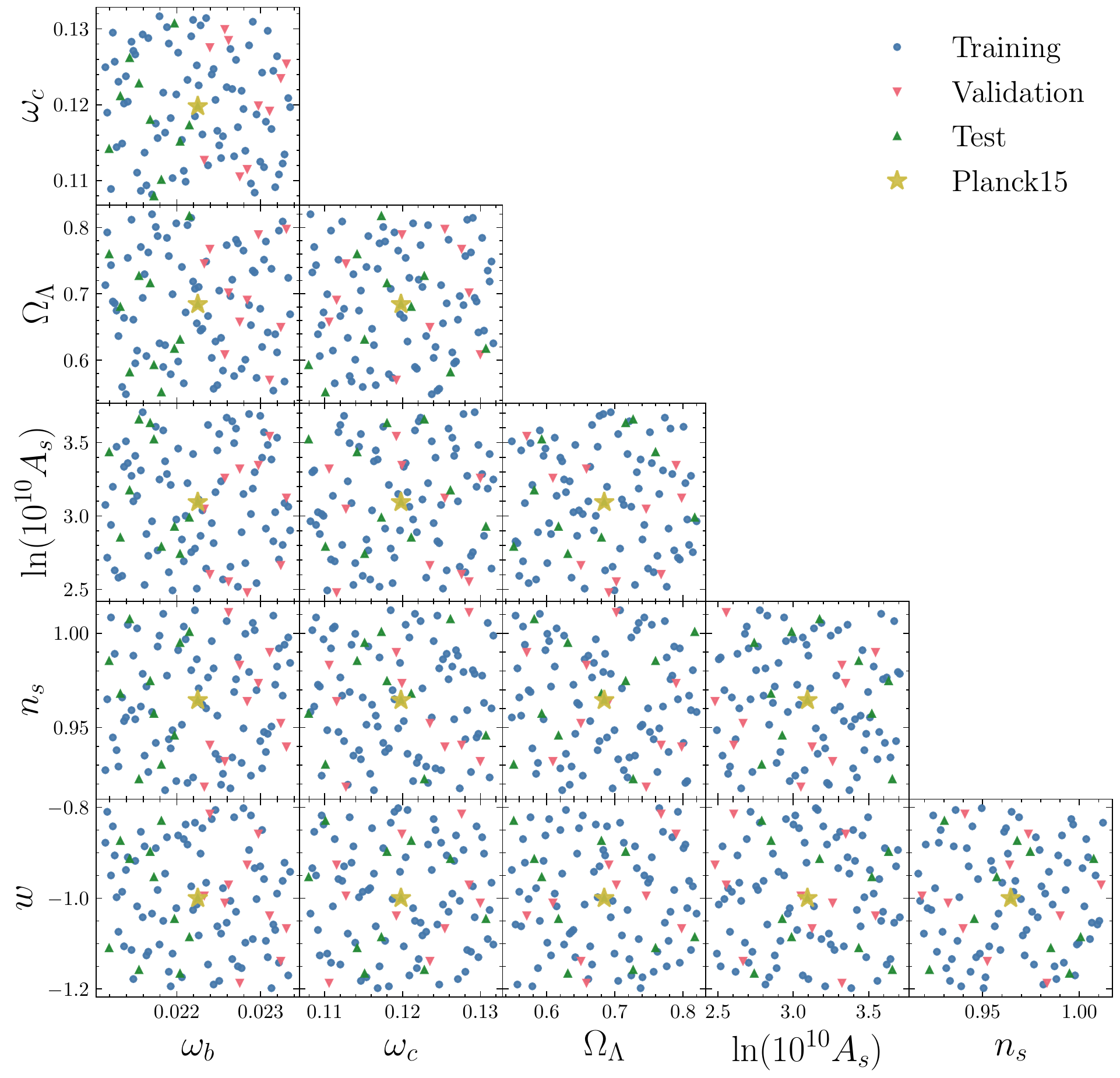}
    \caption{Corner plot representation of the $101$ $w$CDM cosmologies covered by the \textsc{Dark Quest} simulation suite. We show the cosmologies chosen as training, test and validation sets, together with the best fitting fiducial cosmology to the 2015 Planck data, using different symbols, as indicated by the key.} 
    \label{fig:parameter_space}
\end{figure*}

The cosmologies used in the simulations cover $101$ flat geometry $w$CDM models, as shown in Fig.~\ref{fig:parameter_space}. In $w$CDM, the equation of state (EoS) for dark energy is parameterised through the value of $w$, also known as the EoS parameter of dark energy, $p_\mathrm{de} = w \rho_\mathrm{de}$, whose value is $w=-1$ in $\Lambda$CDM. Here, $w$ is assumed to be constant.  

The set of cosmological parameters is defined using optimal maximin distance sliced Latin hypercube designs \citep{doi:10.1080/00401706.2014.957867}, which enable efficient sampling from the six-dimensional parameter space, 
\begin{equation}
    \label{eq:cosmology}
    \mathcal{C} = \qty{\omega_{\rm b}, \omega_{\rm c}, \Omega_{\mathrm{de}}, \ln(10^{10}A_{\rm s}), n_{\rm s}, w},
\end{equation}
where $\omega_{\rm b} \equiv \Omega_{\rm b} h^2$ and $\omega_{\rm c} \equiv \Omega_{\rm c} h^2$ are the physical density parameters of baryons and cold dark matter, respectively. 
The total matter density is the summation of the contributions from baryons, cold dark matter, and non-relativistic neutrinos:
\begin{align}
    \Omega_{\rm m} = \Omega_{\rm b} + \Omega_{\rm c} + \Omega_{\nu},
\end{align}
where the physical density of neutrinos is fixed in the \textsc{Dark Quest} simulations as $\omega_{\nu} \equiv \Omega_{\nu}h^2 \equiv 0.00064$, corresponding to $0.06\,\mathrm{eV}$ for the total mass of the three mass eigenstates.
For given values of $\omega_{\rm b}, \omega_{\rm c}$ and the density parameter for dark energy $\Omega_{\mathrm{de}}$, the Hubble constant is derived from spatial flatness, that is,
\begin{align}
    \Omega_{\rm m} h^2 = \omega_{\rm b} + \omega_{\rm c} + \omega_{\nu}, \\
    \Omega_{\rm m} + \Omega_{\mathrm{de}} = 1. 
\end{align}
$A_{\rm s}$ and $n_{\rm s}$ are the amplitude and slope of the primordial curvature power spectrum normalised at $0.05\,\mathrm{Mpc}^{-1}$.
The range of parameters explored is
\begin{align}
    &0.0211375 < \omega_{\rm b} < 0.0233625, \notag \\
    &0.10782 < \omega_{\rm c} < 0.13178, \notag \\
    &0.54752 < \Omega_{\mathrm{de}} < 0.82128, \notag \\
    &2.4752 < \ln(10^{10}A_{\rm s}) < 3.7128, \notag \\
    &0.916275 < n_{\rm s} < 1.012725, \notag \\
    &-1.2 < w < -0.8, \label{eqn:dq_param_range}
\end{align}
which is centred on the fiducial best fitting $\Lambda$CDM model to the Planck 2015 data alone \citep{Planck15Parameters:2016A&A...594A..13P}: $\omega_{\rm b} = 0.02225, \omega_{\rm c} = 0.1198, \Omega_{\mathrm{de}} = 0.6844, \ln{(10^{10}A_{\rm s})} = 3.094, n_{\rm s} = 0.9645$ and $w=-1$. Fig.~\ref{fig:parameter_space} shows a two-dimensional representation of the parameter space. 

These parameter ranges correspond to the ranges of $(\pm5\%, \pm10\%, \pm20\%, \pm20\%, \pm5\%)$ for the parameters $(\omega_{\rm b}, \omega_{\rm c}, \Omega_{\rm de}, \ln(10^{10}A_{\rm s}), n_{\rm s})$, respectively. These ranges were chosen to cover a parameter space that extends well beyond the constraints from the 2015 Planck data for a flat-$\Lambda$CDM model, for which the corresponding $68\%$ intervals are $(0.72\%, 1.25\%, 1.33\%, 1.10\%, 0.51\%)$. Therefore, the \textsc{Dark Quest}  simulations cover roughly up to a $\sim 10\,\sigma$ range around the central best-fitting model to the Planck 2015 data. However, for the dark energy EoS parameter, $w$, a different approach was taken. Since Planck data alone cannot place a stringent constraint on $w$, and also, assuming that $w$CDM significantly loosens the constraints on the other parameters, we chose a strategy that is not strictly consistent for the six parameters. Instead, we used the Planck data combined with other external data sets only in the case of $w$ (ie, $w=-1.019^{+0.075}_{-0.08}$ at 95\% CL), and tried to cover a much wider range.

The simulation outputs were stored at $21$ redshifts: 1.48, 1.35, 1.23, 1.12, 1.02, 0.932, 0.846, 0.765, 0.689, 0.617, 0.549, 0.484, 0.422, 0.363, 0.306, 0.251, 0.198, 0.147, 0.0967, 0.0478, and 0.
These redshifts are evenly spaced in the linear growth factor for the fiducial Planck cosmology.
\subsection{Halo catalogues}
\label{subsec:halo_catalogs}
The identification of halos is of crucial importance, since the central premise of our method is to emulate dark matter halo properties, which can be robustly measured from $N$-body simulations.
Appendix~E of the \textsc{Dark Quest} paper \citep{Nishimichi:2018etk} provides comprehensive convergence tests of halo properties such as halo mass, the halo mass functions, and halo autocorrelation functions, with respect to the choice of halo finder, halo substructure separation, central/satellite split criterion, etc. In this section, we briefly review the main definitions that will be used in this paper.

The halo catalogues used here were identified using \textsc{Rockstar} \citep{Behroozi:2013ApJ...762..109B}, a friends-of-friends (FOF) halo finder that operates in six-dimensional phase space.
The halo centre is defined as the centre of mass position of the ``core particles'', a subset of member particles in the inner part of the halo.
$M_{200m}$ is adopted as the halo mass definition in \textsc{Dark Quest}, which is the mass enclosed within $R_{200m}$, the radius within which the average density is $200$ times the mean mass density $\bar{\rho}_{m0}$.
This definition of halo mass includes all simulation particles within a radius of $R_{200m}$ from the halo centre, including gravitationally unbound ones. 
When the separation between the centres of different halos is within $R_{200m}$ of any other halo, the most massive halo is marked as a central halo and the other halo(s) as a satellite halo(s). 
Only central halos with mass $M_{200m} \ge 10^{12} \, h^{-1}M_{\odot}$ are used in our analysis.

\section{From dark matter halos to galaxies}
\label{sec:halo_to_galaxy}
As in \citet{Nishimichi:2018etk}, \citet{2021arXiv210100113M} and \citet{PhysRevD.102.063504} we use the halo model to express the galaxy two-point correlation function in terms of dark matter halo properties. This allows us to make theoretical predictions for different galaxy samples, including cross-correlations of two different tracers, such as the ones that would be used in a multitracer analysis \citep{McDonald_2009}, or the cross-correlation between clusters and galaxies. Moreover, a halo model implementation allows us to model the halo-galaxy connection analytically, which means that the accuracy of the results will not be worsened by emulator inaccuracies. As a downside, complex models of the halo-galaxy connection such as environment-based assembly bias may be harder to implement. 

The halo model assumes that galaxies occupy dark matter halos, and therefore that the two-point galaxy  correlation function can be split into contributions from galaxy pairs  that inhabit the same dark matter halo, and pairs in which each member  occupies a different dark mater halo (these terms will be referred to as the one and two halo terms, respectively):
\begin{equation}
    \xi_{\rmgg}(r) = \xi_{\rmgg}^{\mathrm{1h}}(r) +  \xi_{\rmgg}^{\mathrm{2h}} (r).
\end{equation}
The one and two halo terms can be further split into correlations between two types of galaxies: centrals and satellites. Central galaxies are positioned at the minimum of the potential well of the dark matter halo and move with the halo's centre of mass velocity. Satellite galaxies orbit within the dark matter halo with virialised velocities. We assume that the distribution of satellite galaxies is given by an NFW profile, $u_{\mathrm{NFW}} (r|c(M))$ \citep{1997ApJ...490..493N}. This approximation has been tested against hydrodynamical simulations, finding it valid for galaxies selected by number density \citep{2019MNRAS.490.5693B}. The NFW profile is defined by one parameter: the concentration of the halo, $c$, which varies with halo mass, redshift, and cosmological parameters \citep{2016MNRAS.460.1214L, Diemer_2019}. Here, we use the median concentration-mass relation $c(M)$ from \citet{Diemer_2019}.

Regarding the galaxy-halo connection, we use the halo occupation distribution (HOD) \citep{2005ApJ...633..791Z} to model the number of galaxies in a given halo as a function of halo mass. The occupation of central galaxies is parameterized as a Bernoulli distribution, whereas that of satellites is assumed to be Poisson distributed.  Both distributions are described by their mean parameters
\begin{equation}
    \left\langle  N_{\rmg} \right\rangle (M) =  \left\langle  N_{\mathrmc} \right\rangle(M) + \left\langle  N_{\mathrms} \right\rangle(M).
\end{equation}
We parameterize the mean galaxy numbers as in \citet{2005ApJ...633..791Z} by introducing the following HOD parameters
\begin{equation}
    \mathcal{G} = \{ M_\mathrm{min}, \sigma_{\log M}, M_1, \kappa, \alpha \},
\end{equation}
where $M_\mathrm{min}, \sigma_{\log M} $, and $ M_1, \kappa, \alpha$ define the occupation of the centrals and satellites, respectively.

We describe the mean number of central galaxies for a given halo as
\begin{equation}
    \left\langle  N_{\mathrmc} \right\rangle(M|\mathcal{G}) = \frac{1}{2} \left( 1 + \erf \left(\frac{\log M - \log M_{\mathrm{min}}}{\sigma_{\log M}} \right) \right),
    \label{eq:occ_centrals}
\end{equation}
where $\erf(x)$ is the error function. The mean occupation number of satellite galaxies is defined as
\begin{align}
 \left\langle  N_{\mathrms} \right\rangle(M|\mathcal{G}) &= \left\langle  N_{\mathrmc} \right\rangle(M|\mathcal{G}) \, \lambda_{\mathrms} (M|\mathcal{G}) \notag \\
 &= \left\langle  N_{\mathrmc} \right\rangle(M) \left(\frac{M - \kappa M_{\mathrm{min}} }{M_1} \right)^\alpha.
     \label{eq:occ_satellites}
\end{align}

The empirical HOD model that we use is extremely simple. One of the simplifying assumptions is that galaxy occupation depends solely on the mass of the dark matter halo. Although dark matter halo mass correlates strongly with clustering, we know that dark matter halos experience different assembly histories even at a fixed halo mass, which can affect their clustering \citep{Gao:2005MNRAS.363L..66G.AssembleBias,Gao:2007MNRAS.377L...5G.AssembleBias}.
These different assembly histories influence secondary properties of halos, and this might, in turn, affect the formation of galaxies and hence the galactic content of halos of a given mass. These effects together -- the variations in halo clustering and galactic content with halo mass and a second halo property --  are known as galaxy assembly bias (see \citealt{2018ARA&A..56..435W} for a recent review on the galaxy-halo connection and assembly bias). The question we will address in Section~\ref{sec:ab}, is whether a simplified version of the galaxy-halo connection is flexible enough to recover unbiased constraints on the cosmological parameters.

Given these assumptions, we can express the two-point galaxy correlation function in terms of dark matter halo properties. To simplify the calculations, we further split the one and two halo terms into correlations of central and satellite galaxies,
\begin{equation}
    \xi_{\rm gg}(r) = \xi_{\rm ss}^{\rm 1h}(r) + 2 \xi_{\rm cs}^{\rm 1h}(r)  + \xi_{\rm cc}^{\rm 2h}(r)  + 2 \xi_{\rm cs}^{\rm 2h}(r)  + \xi_{\rm ss}^{\rm 2h}(r).
    \label{eq:halo_model}
\end{equation}
In the equations below, we highlight the emulated quantities in \textcolor{blue}{blue}, such as the halo mass functions, {\color{blue}$\displaystyle {\rm d}{n}/{\rm d}{M}$}, and halo auto correlation functions, {\color{blue}$\xi_{ {\rm hh}}(r)$}, following the convention used in \citet{2021arXiv210100113M}.  Note that terms involving both centrals and satellites lead to the convolution of the halo profiles and the halo two-point correlation function. It is therefore simpler to compute these terms in Fourier space, where convolutions in coordinate space become simple products, and then apply an inverse Fourier transform to the result. Therefore, we compute
\begin{equation}\label{eq:ss_1h}
    P_{\rm ss}^{\rm 1h}(k) = \frac{1}{\bar{n}_{\rm g}^2} \int {\rm d} M \textcolor{blue}{\frac{ {\rm d} n}{ {\rm d}M}(M)} \left\langle  N_{\rm c} \right\rangle(M)\lambda^2_{\rm s}(M) u_{\mathrm{NFW}} (k|M, c(M))^2,
\end{equation}
where $u_{\mathrm{NFW}} (k|M, c(M))$ is the Fourier transform of the truncated NFW profile (see Eq.~(81) in \citealt{COORAY_2002}). 

The cross-correlation between centrals and satellites that occupy the same halo is given by
\begin{equation}
    \label{eq:cs_1h}
    P_{\rm cs}^{\rm 1h}(k) = \frac{1}{\bar{n}_{\rm g}^2} \int {\rm d} M \textcolor{blue}{\frac{ {\rm d} n}{{\rm d}M}(M)} \left\langle  N_{\rm c} \right\rangle(M)\lambda_{\rm s}(M) u_{\mathrm{NFW}} (k|M, c(M)),
\end{equation}
where ${\rm d} n/ {\rm d}M (M)$ is the halo mass function defined as the comoving number density of halos for a given halo mass, and $\bar{n}_{\rm g}$ is the galaxy number density that we obtain by integrating the halo mass function weighted by the halo occupation
\begin{equation}
\label{eq:n_g}
    \bar{n}_{\rm g} =\int {\rm d}M \textcolor{blue}{\frac{ {\rm d} n}{ {\rm d}M}} \left( \left\langle N_{\rm c} \right\rangle (M) + \left\langle N_{\rm s} \right\rangle (M) \right).
\end{equation}

Meanwhile, the different two-halo terms will result in weighted averages of the dark matter halo two point correlation function and convolutions with NFW profiles when satellite correlators are involved
\begin{equation}
    \label{eq:cs_2h}
    \begin{split}
    P_{\rm cs}^{\rm 2h}(k) = & \frac{1}{\bar{n}_{\rmg}^2} \int {\rm d} M \textcolor{blue}{\frac{ {\rm d} n}{{\rm d}M}(M)} \left\langle  N_{\rm c} \right\rangle(M) \\ 
    & \int {\rm d} M^\prime \textcolor{blue}{\frac{ {\rm d} n}{ {\rm d}M}(M^\prime)} \left\langle  N_{\rm c} \right\rangle(M^\prime) \lambda_{\rm s}(M^\prime)  \\ & \textcolor{blue}{P_{\rm hh}(k|M, M^\prime)} u_{\mathrm{NFW}} (k|c(M^\prime)),
    \end{split}
\end{equation}

\begin{equation}
    \label{eq:ss_2h}
    \begin{split}
    P_{\rm ss}^{\rm 2h}(k) = & \frac{1}{\bar{n}_{\rmg}^2} \int {\rm d} M \textcolor{blue}{\frac{ {\rm d} n}{{\rm d}M}(M)} \left\langle  N_{\rm c} \right\rangle(M) \lambda_{\rm s}(M) \\ & \int {\rm d} M^\prime \textcolor{blue}{\frac{ {\rm d} n}{ {\rm d}M}(M^\prime)} \left\langle  N_{\rm c} \right\rangle(M^\prime) \lambda_{\rm s}(M^\prime) \\ & \textcolor{blue}{P_{\rm hh}(k|M, M^\prime)} u_{\mathrm{NFW}}(k|c(M^\prime)) u_{\mathrm{NFW}}(k|c(M)).
    \end{split}
\end{equation}

We avoid the Fourier transform when computing central-central terms
\begin{equation}
\begin{split}
    \label{eq:cc_2h}
    \xi_{\rm cc}^{\rm 2h}(r) = & \frac{1}{\bar{n}_{\rm g}^2} \int {\rm d} M \textcolor{blue}{\frac{ {\rm d} n}{ {\rm d}M}(M)} \left\langle  N_{\rm c} \right\rangle(M) \\ & \int {\rm d} M^\prime \textcolor{blue}{\frac{ {\rm d} n}{ {\rm d}M}(M^\prime)} \left\langle  N_{\rm c} \right\rangle(M^\prime) \textcolor{blue}{\xi_{\rm hh}(r|M, M^\prime)}.
\end{split}
\end{equation}
In the next section, we show how we can use neural networks to emulate the two statistics shown in blue that vary with cosmological parameters: ${\rm d} n / {\rm d}M$ and $\xi_{\rm hh}$.

\subsection{The best of both universes: combining simulations of different resolutions}
\label{sec:hr_lr}
Although the high-resolution (HR) simulations can resolve halos of lower masses than their low-resolution (LR) counterparts, their smaller box size results in a larger sample-variance noise than in the LR boxes.

The halo model approach outlined above allows us to calibrate the halo autocorrelation function using the LR simulations, to reduce sample variance when using measurements from one realisation, while calibrating the halo mass function with the HR simulations to ensure an accurate estimate of the halo mass function for low mass halos. In this section, we examine the impact of combining the halo mass function of HR simulations with the halo correlation function measured in LR simulations.\footnote{Note we could also have extended the mass resolution of the LR halo catalogues, using a scheme like the introduced by \citet{Joaquin:2022}.}

\begin{figure}
    \centering
    \includegraphics[width=0.45\textwidth]{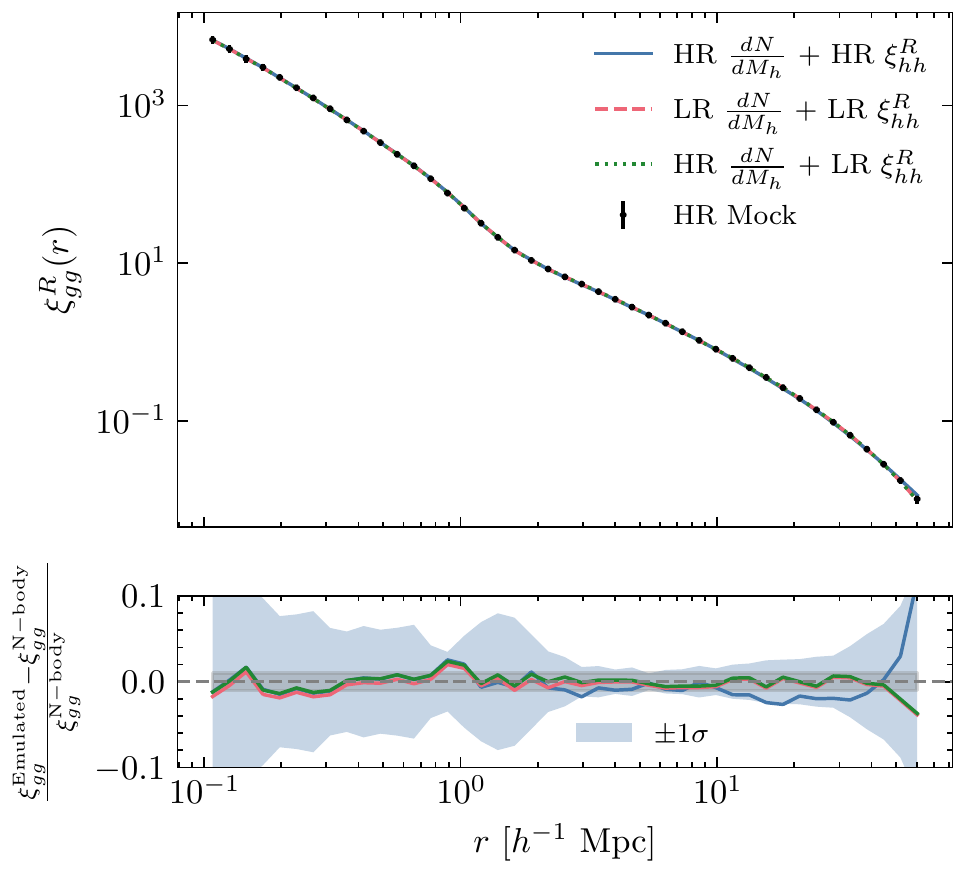}
    \caption{We show $\xi_{\rm gg}^R$ obtained by populating the $25$ realizations of the fiducial cosmology on the HR simulations with mock LOWZ galaxies, compared to the result of Eq.~\ref{eq:halo_model} when either: i) both ${\rm d}N/{\rm d}M_{\rm h}$ and $\xi_{\rm hh}^R$ are measured on the HR simulations (in blue), ii) both ${\rm d}n/ {\rm d}M_{\rm h}$ and $\xi_{\rm hh}^R$ are measured on the LR simulations (in red) and iii) ${\rm d}n/{\rm d}M_{\rm h}$ is obtained from the HR simulations and $\xi_{\rm hh}^R$ from the larger boxsize LR ones (in green). The fractional difference plot in the lower panel shows that the sample variance in the blue line based on the correlation function measured from one HR box is greatly reduced by replacing it with LR simulations without introducing bias. Blue shaded denote the standard deviation of the $25$ realizations of the HR simulations. the gray shaded regions denotes $1\%$ errors.} 
    \label{fig:combined}
\end{figure}

In Fig.~\ref{fig:combined}, we show a comparison of a mock LOWZ-like catalogue obtained from the $25$ realisations of the fiducial cosmology for the HR simulations, to the result of Eq.~\ref{eq:halo_model} when i) we combine the halo mass function from HR simulations, with the halo two-point correlation function estimated from one of the HR boxes (solid blue line), ii) estimate both the halo mass function and halo two-point correlation function from the LR simulations (dashed red), and iii) measure the halo mass function in the HR simulation, and the halo auto-correlation from the LR simulation. Fig.~\ref{fig:combined} shows that combining clustering measurements from low-resolution simulations with a halo mass function measured in the HR simulation does not introduce any biases and reduces the sample-variance noise.

\section{Neural Network emulators for dark matter halo properties}
\label{sec:neural_nets}
\citet{Nishimichi:2018etk} fitted both the halo mass function and the halo autocorrelation function measured from the N-body simulations using a combination of principal component analysis (PCA), to reduce the dimensionality of the data vector, and Gaussian processes (GP), to fit the dependence of the principal component coefficients on cosmology. Here, we show how dimensionality reduction can be avoided by using neural network emulators, leading to increased accuracy in the prediction of halo properties.

Fully connected neural networks approximate a function $f$ such that
\begin{equation}
   \mathbf{y} = f(\mathbf{x}|\boldsymbol{\theta}), 
\end{equation}
where $\mathbf{x}$ represents the features of the data set, $\mathbf{y}$ the desired outputs, and $\boldsymbol{\theta} $ the network-free parameters, also called trainable parameters. The optimal function $f$ is defined by the set of values $\theta$ that minimise the loss function (the form of which is discussed below). The loss function provides a measure of the model's performance when evaluated on the data set.

 ReLU (Rectified Linear Unit; \citealt{agarap2018deep}) is the most commonly used activation function in current neural networks used to add non-linearities in the mapping between inputs and outputs, and is defined as
 \begin{equation}
     \mathrm{ReLU}(x) = \max(0,x),
 \end{equation}
 where $x$ is the output of the previous layer of the neural network. Note that ReLU activations are not differentiable at zero. Here, however, we are interested in functions that are differentiable with respect to their inputs and, in particular, with respect to the cosmological parameters (since these derivatives could be used to accelerate parameter inference through Hamiltonian Monte Carlo techniques, e.g. \citealt{Duane:1987de}, or to accelerate Fisher forecasts). Therefore, throughout, we use Gaussian error linear units (GELUs) as activation functions instead \citep{2016arXiv160608415H}:
 \begin{equation}
     \mathrm{GELU}(x) = 0.5x\left(1 + \erf\left(\frac{x}{\sqrt{2}}\right)\right).
 \end{equation}

To find the optimal parameters, $\theta$, which reproduce the statistics measured from the N-body simulations, we minimise the L1 norm loss function
\begin{equation}
    \label{eq:loss}
    \mathcal{L} = \frac{1}{N} \sum_{i=0}^N |y^i_\mathrm{true} - y^i_\mathrm{predicted}|,
\end{equation}
using the Adam optimiser \citep{2014arXiv1412.6980K}.  Note that L1 reduces the importance given to outlier errors compared to the use of the mean squared error (also known as the L2 norm). We will refer to the value of Eq.~\ref{eq:loss} evaluated in the training and validation dataset as training and validation loss, respectively.

Moreover, we avoid fine-tuning the value of the learning rate by using a learning rate scheduler that reduces the learning rate by a factor of $10$ every time the validation loss does not improve after $20$ epochs. We also stop training the model when the validation loss does not improve after $100$ epochs. This iterative reduction of the learning rate allows the model to quickly learn the broad characteristics of the data and then reduce the errors by adopting a smaller learning rate. The initial learning rate is always set to $0.015$.



In the following subsections, we demonstrate the precision of fully connected networks in reproducing the real-space correlation function and the halo mass function obtained from the \textsc{dark quest} simulations.

\subsection{Real space correlation function}
\subsubsection{Measurement}
The details of the halo correlation function measurements are introduced in \citet{Nishimichi:2018etk}. Here, we present only a summary of the most important aspects. 

First, noisy measurements of $\xi(r|M, M^\prime)$ are avoided by instead measuring $\xi$ as a function of halo number density, $n$, and switching from differential to cumulative mass limits. 
We then use the halo mass function to translate predictions as a function of number density into predictions as a function of differential mass through the relation
\begin{equation}
\begin{split}
    \xi(r| n(m), n(m^\prime)) & = \frac{\int_m^\infty {\rm d}M \int_{m^\prime}^\infty  {\rm d}M^\prime \xi(r|M, M^\prime) \frac{ {\rm d}n}{{\rm d}M}(M) \frac{ {\rm d}n}{{\rm d}M}(M^\prime) }{\int_m^\infty {\rm  d}M \int_{m^\prime}^\infty  {\rm d}M^\prime \frac{ {\rm d}n}{{\rm d}M}(M) \frac{ {\rm d}n}{ {\rm d}M}(M^\prime) } \\
    & = \frac{\int_m^\infty {\rm d}M \int_{m^\prime}^\infty  {\rm d}M^\prime \xi(r|M, M^\prime) \frac{ {\rm d}n}{ {\rm d}M}(M) \frac{ {\rm d}n}{{\rm d}M}(M^\prime) }{n(M) n(M^\prime)},
\end{split}
\end{equation}
which can be inverted to obtain
\begin{equation}
\label{eq:xi_conversion}
\begin{split}
    \xi(r|M, M^\prime) & = \frac{\frac{\partial^2}{\partial m \partial m^\prime} \left[n(m) n(m^\prime) \xi(r|n(m), n(m^\prime))\right]}{ \frac{ {\rm d}n}{{\rm d}M}(M) \frac{{\rm d}n}{ {\rm d}M}(M^\prime) } \\
    & = \frac{\partial^2}{\partial n \partial n^\prime}\left[n(m) n(m^\prime) \xi(r|n(m) n(m^\prime))\right].
\end{split}
\end{equation}

Measurements are made in $8$ logarithmically spaced bins in number density over the range $n_{\rm h} = \left[ 10^{-6}, 10^{-2.5}\right]$ $\left(h^{-1}\mathrm{Mpc}\right)^{-3}$. Note that there are $36$ independent combinations for two halo samples with different number densities. The pair separation $r$ is split into $40$ logarithmically spaced bins from $0.01$ to $5$ $h^{-1}\,\mathrm{Mpc}$ and $75$ linear bins from $5$ to $150$ $h^{-1}\,\mathrm{Mpc}$, and over the $21$ simulation snapshots spanning from $z=1.48$ to $z=0$.

In total, the data set is made up of $80$ cosmologies in the training set, $10$ in the validation set and $10$ in the test set, each with its corresponding $21$ snapshots and $36$ number density bins.

On large scales, we can reduce cosmic variance by using the propagator-based prescription of \citet{2006PhRvD..73f3520C}. For Gaussian initial conditions, the propagator can be expressed as the ratio of the cross-power spectrum between the density field at the initial conditions and the nonlinear field at the redshift of interest,  to the linear power spectrum. This calculation was originally performed for the matter density, but can be extended to the halo density field. The propagator  quantifies how much of the memory of the initial conditions is preserved in the final nonlinear density field. The propagator  describes the smearing of BAO feature due to large-scale bulk flows. One can straightforwardly generalize this approach to any tracer. This function also describes the linear bias factor in the large-scale limit. The advantage of using the propagator is that a large fraction of sample-variance error is cancelled when the ratio between the two spectra is taken. In addition, it is known that the $k$ dependence of the propagator is simple. A Gaussian-like parameterized function is sufficient to model this accurately (see \citealt{Nishimichi:2018etk} for more details).

We have slightly updated the implementation of this idea here. In \cite{Nishimichi:2018etk}, to evaluate the correlation function, both the directly emulated correlation function (for small separations) and the propagator-based model (for large separations), in which the propagator is also emulated, are computed and then stitched together to cover a wide range of separations. This requires us to build two separate emulators and both of them must be used when evaluating the correlation function. Here, instead, we now work at the data level: for each simulation box, we construct a data vector that combines the two methods. We refined the stitching scheme to yield  a smoother  transition between the two regimes (Nishimichi et~al. in prep.). Now, our neural-network emulator learns this new datavector, to which the propagator trick has already been applied.

\subsubsection{Emulation}
We train a fully connected neural network, $f$, to perform the following mapping
\begin{equation}
   \log_{10}\left( \xi^{\mathrm{R}}_{\rmhh}(r)\right) = f(\mathcal{C}, \log_{10}(n_1), \log_{10}(n_2), z),
\end{equation}
where $n_1$ and $n_2$ denote the number densities of each halo sample, $z$ is the redshift and $\mathcal{C}$ represents the set of cosmological parameters in Eq.~\ref{eq:cosmology}.

Note that the input to the neural network has been standardised to facilitate training (such that its mean is $0$ and standard deviation is $1$). The output of the neural network is the logarithm of the correlation function $\log_{10}(\xi_{\rmhh})$, which is also standardised:
\begin{equation}
    \log_{10} \qty(\xi^{\mathrmR}_{\rmhh}(r)) \rightarrow \frac{ \log_{10} \qty(\xi^{\mathrmR}_{\rmhh}(r)) - \left\langle \log_{10} \qty(\xi^{\mathrmR}_{\rmhh}(r)) \right\rangle }{\sqrt{\mathrm{Var} \qty(\log_{10} \qty(\xi^{\mathrmR}_{\rmhh}(r) ))}},
\end{equation}
where $\left\langle \log_{10} \qty(\xi^{\mathrmR}_{\rmgg}(r)) \right\rangle$ and $\mathrm{Var} \qty(\log_{10} \qty(\xi^{\mathrmR}_{\rmgg}(r) ))$ are the mean and variance of all correlation functions, estimated from the training set.

The output of the neural network is all the values of the correlation function evaluated for the pair-separation vector, $r$. Interestingly, when fitting the neural network with $r$ as input, the model tends to overfit the data and converges to a less accurate overall model, while combining all pair separations shares the weights of the neural network across the values of $r$ and reduces the level of overfitting.

We summarise the best-fitting hyperparameters of the neural network in Table~\ref{table:hyperparameters}. 

\begin{figure*}
    \centering
    \includegraphics[width=0.7\textwidth]{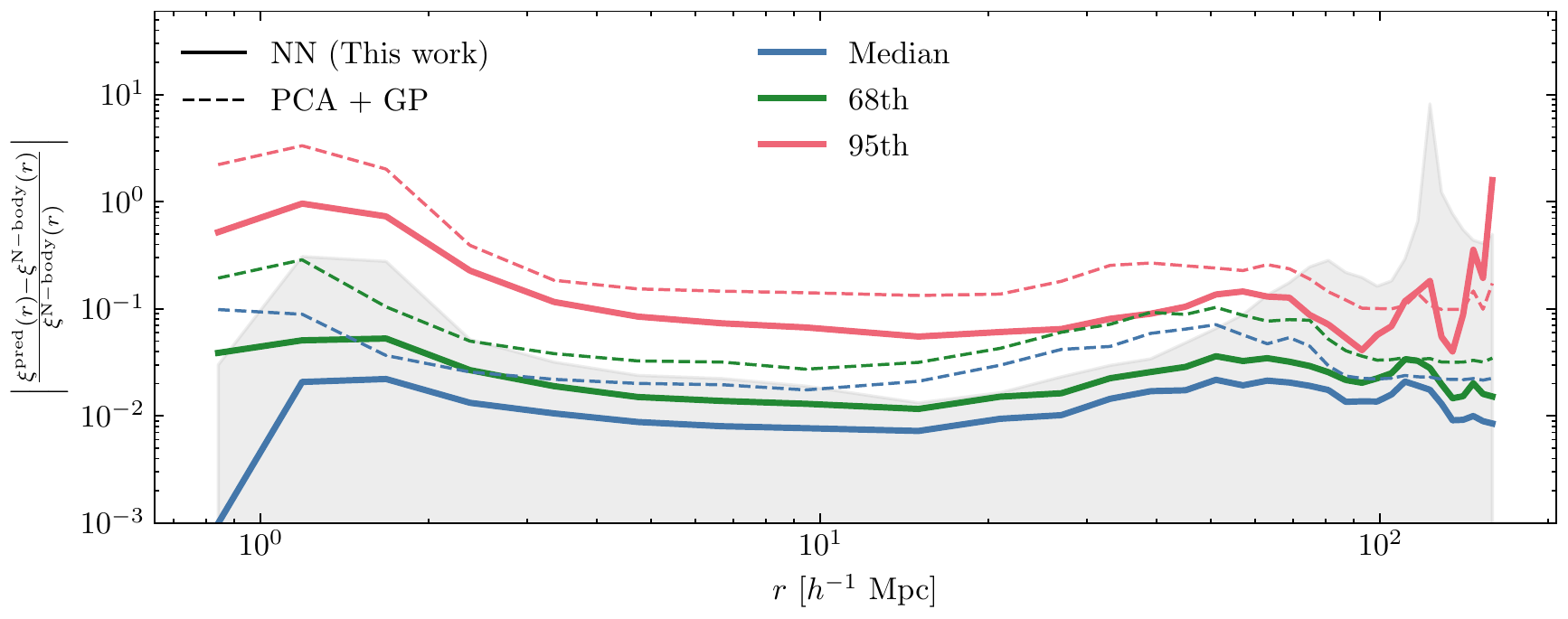}
    \caption{Comparison of the absolute fractional errors of the neural network emulator for the halo real space two point correlation function, with the Gaussian process + PCA approach presented in \citet{Nishimichi:2018etk}. Note that we only include test set data, but for all redshifts and halo number densities. The grey shading shows the variance estimated from the simulations using the $15$ realisations of the fiducial Planck cosmology, $\sigma_{\xi_\mathrm{fiducial}}/\xi_\mathrm{fiducial}$.}
    \label{fig:xi_hh_performance}
\end{figure*}

In Fig.~\ref{fig:xi_hh_performance}, we show the performance of the neural network as a function of pair separation compared to that found in \citet{Nishimichi:2018etk}. Fig.~\ref{fig:xi_hh_performance} shows the absolute errors estimated in the test set, as a function of pair separation $r$. Number densities and redshifts have been averaged. 

The median absolute errors are lower than $2\%$ throughout the entire scale range, a factor of $4$ smaller than the upper limit of \citet{Nishimichi:2018etk}, while $68\%$ had errors smaller than $6\%$, which is a factor of $5$ smaller. We further compare the variance of the emulator errors (68th percentile fractional residuals) to the variance in the simulations themselves (grey solid background). This comparison shows that the emulator is already performing at a level similar to the variance in the simulations over the full-scale range. Note also that we cannot accurately estimate the model accuracy below the level of 
sample
variance in the simulations, given that we only compare the accuracy of the model against one N-body realisation for each cosmology in the test set.

\begin{table*}
\centering
\caption{The summary of the best performing set of hyperparameters for the neural network emulators used to predict halo properties. The last column indicates the simulation resolution from which the quantity listed in the first column is measured.}
\label{table:hyperparameters}
\begin{tabular}{c c c c c}
\hline
\hline
Statistic & Batch size & Activation &  $N_\mathrm{hidden}$ & Resolution\\ 
\hline
\hline
$\displaystyle\xi_{\rmhh}$ & $5000$ & GELU &  $1024, \, 512, \, 512$ & LR \\
\hline
$\displaystyle \dv{n}{M}$ & $5000$ & GELU & $1024, \, 512, \, 512$ & HR \\
\hline
\end{tabular}
\end{table*}

\subsection{Halo mass function}
\subsubsection{Measurement}
As explained earlier, we used the HR simulations to model the halo mass function. To do this, we first create a histogram of the number of halos in 80 logarithmically spaced bins in halo mass over the range of $10^{12}$ to $10^{16}\,h^{-1}\,M_\odot$. Following \cite{Nishimichi:2018etk}, we apply a correction to individual halo masses to account for systematics due to the finite number of particles. The corrected mass is given by (e.g. \citealt{Warren:2006}):
\begin{eqnarray}
\tilde{M} = (1+N_\mathrm{p}^{-0.55})M,
\end{eqnarray}
where $N_\mathrm{p}$ is the number of simulation particles contained in the halo.
The raw histogram is rather noisy, especially at the high-mass tail due to the small number of halos per bin. To produce a smooth mass function, we fit the data points using the functional form employed in \cite{Tinker2008}. In doing so, we fix the parameter ``$b$'' in the formula, which controls the low mass behaviour, to the original value in \cite{Tinker2008} and allow the other three parameters to vary freely. We weight the bins according to the Poisson noise, which is more important at high masses, and the mass-determination accuracy, which is sensitive to the number of particles in the halo
\begin{eqnarray}
\frac{\Delta N_\mathrm{h}}{N_\mathrm{h}} = \frac{1}{\sqrt{N_\mathrm{h}}}+\frac{1}{N_\mathrm{p}}.
\end{eqnarray}
The uncertanties in the fitted parameters are propagated to the smooth model prediction to obtain the expectation value, as well as the uncertanties of the estimated halo number counts in each mass bin.

\subsubsection{Emulation}
As in the case of the halo two-point correlation function, we train the model on the logarithm of the halo mass function to reduce the dynamic range of the observable. In this case, the mapping we obtain is
\begin{equation}
    \log_{10}\left(\dv{n}{M} (M)\right) = f(\mathcal{C}, z).
\end{equation}
As before, we standardise inputs and outputs before training the model.

\begin{figure}
    \centering
    \includegraphics[width=0.47\textwidth]{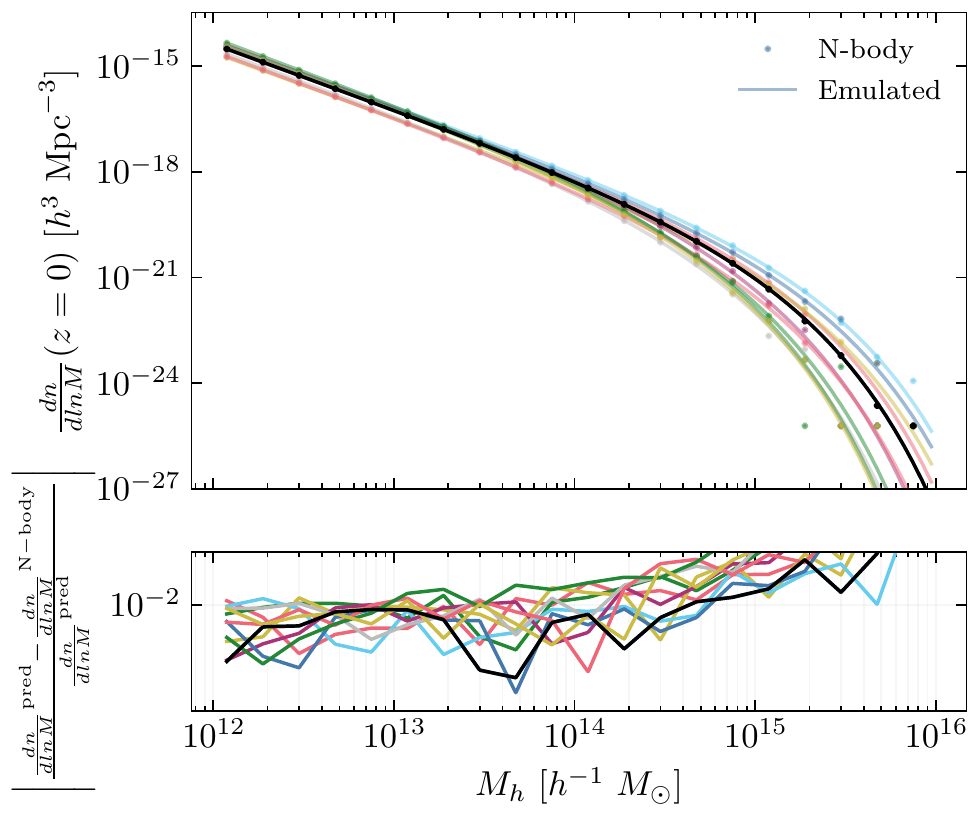}
    \caption{N-body measurements (points) and emulator predictions (lines) for the halo mass function at $z=0$ in the $10$ test set cosmologies. The lower panel shows the absolute fractional errors as a function of halo mass. The fiducial Planck cosmology is shown in black.}
    \label{fig:hmf_fixed_z}
\end{figure}

In Fig.~\ref{fig:hmf_fixed_z}, we compare the N-body measurements from the $10$ test cosmologies with the emulator predictions at $z=0$. The emulator achieves subpercent accuracy for halo masses smaller than $10^{14}$ $h^{-1} \,M_{\odot}$, with the error increasing for larger halo masses. Estimating the error is, however, challenging for halo masses larger than $10^{14}$ $h^{-1} \, M_{\odot}$ due to the large Poisson noise that affects the measuremenents caused by the small number of cluster-size halos in the simulations.

\begin{figure*}
     \centering
     \begin{subfigure}[b]{0.47\textwidth}
         \centering
         \includegraphics[width=\textwidth]{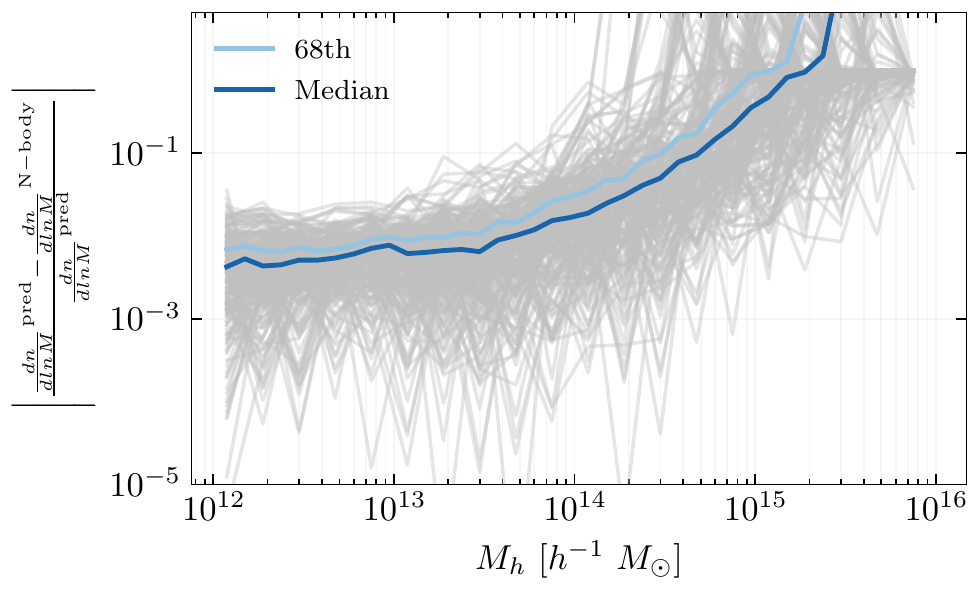}
     \end{subfigure}
     \hfill
     \begin{subfigure}[b]{0.47\textwidth}
         \centering
         \includegraphics[width=\textwidth]{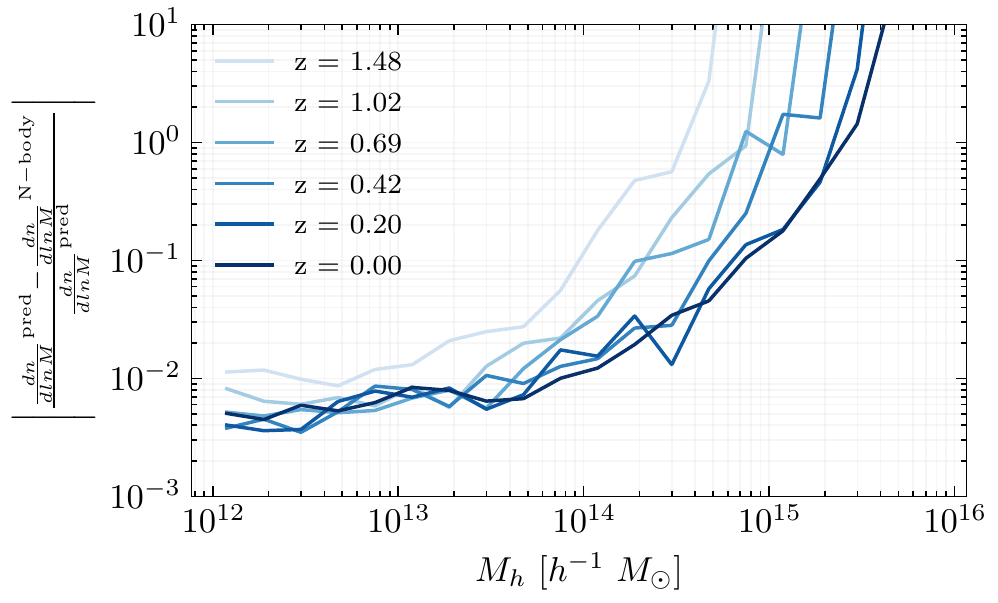}
     \end{subfigure}
     \hfill
    \caption{Absolute fractional errors on the halo mass function emulator predictions as a function of halo mass. The left panel shows the result for each test set sample (the $10$ set cosmologies evaluated at the $21$ different redshifts) as a gray line, along with  the median (dark blue line) and 68th percentile range (light blue line) of the absolute fractional errors. The right panel shows the median absolute error as a function of halo mass, with different lines showing different redshifts, as indicated by the legend.}
    \label{fig:hmf_emu}
\end{figure*}

In Fig.~\ref{fig:hmf_emu}, we evaluate the overall accuracy of the halo mass function emulator at all redshifts (left panel) and as a function of the redshift (right panel). We find that the median emulator error for all redshifts is below $1$ per cent for halo masses smaller than $10^{13.5}\,h^{-1} \,M_{\odot}$, and increases rapidly to values larger than $10$ per cent for the most massive halos ($M_{\mathrm{h}} > 10^{15} \, $ $h^{-1}\, M_{\odot}$). The right panel of Fig.~\ref{fig:hmf_emu} shows that the accuracy of the emulator degrades slightly at the highest redshifts considered ($z=1.48$).

\subsection{Galaxy clustering}

We now assess the impact that inaccuracies in halo emulators have on galaxy clustering predictions. To do so, we populate the $10$ test and $10$ validation LR simulations with mock galaxies. We populate each cosmology at four different snapshots (z=0.1,0.25,0.5 and 0.75) and $5$ different galaxy number densities, logarithmically spaced between $\log \left( \bar{n}_\mathrm{gal}/(h^{-1}\mathrm{Mpc})^{-3}\right) = -3.7$ and $\log \left( \bar{n}_\mathrm{gal}/(h^{-1}\mathrm{Mpc})^{-3}\right) = -4.3$. Note that halo property emulators cannot estimate galaxy clustering for arbitrary number densities, given that the lowest halo mass resolved by the \textsc{Dark Quest} simulations is $10^{12} \, h^{-1} \, M_{\odot}$. 

For each combination of cosmology, redshift, and number density, we randomly sampled the HOD parameters from the ranges
\begin{flalign}
\sigma_{\log{M}} &\in \left[0.1,0.8\right] \nonumber \\
\alpha_{\mathrm{sat}} &\in \left[0.5,1.\right] \nonumber \\
\kappa &\in \left[0.1, 0.8 \right] \nonumber \\
\log M_1 &\in \left[13.5,14.5 \right]. \nonumber
\end{flalign}
The remaining HOD parameter, $\log M_{\mathrm{min}}$, is fixed by the given galaxy number density. In total, we built a diverse sample of $400$ HOD mocks with varying cosmology, HOD parameters, and redshift, to test the performance of the emulator.

\begin{figure}
    \centering
    \includegraphics[width=0.47\textwidth]{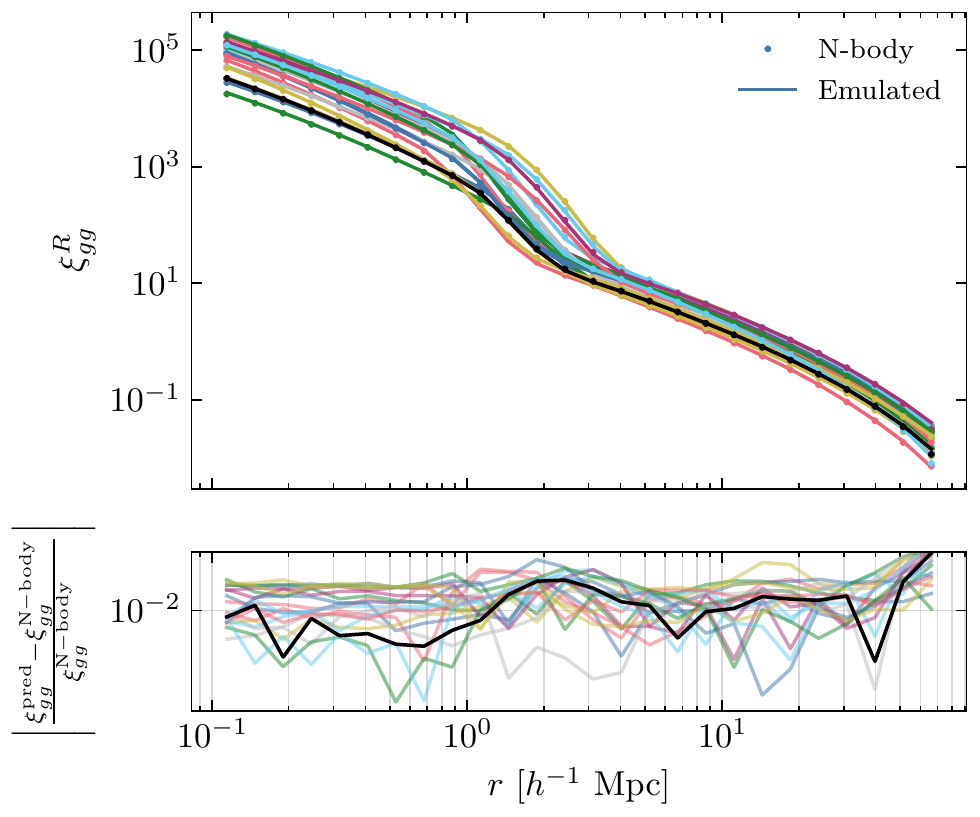}
    \caption{Emulator predictions for a subset of the $400$ HOD mocks generated to test the accuracy of galaxy clustering. We show only those at $z=0.25$. Planck cosmology is shown in black. The top panel shows all measurements from the $20$ HOD catalogues and the corresponding emulator prediction. On the bottom pannel, we show the absolute error of the emulator as a function of scale. } 
    \label{fig:galaxies_eval_z}
\end{figure}

\begin{figure}
    \centering
    \includegraphics[width=0.47\textwidth]{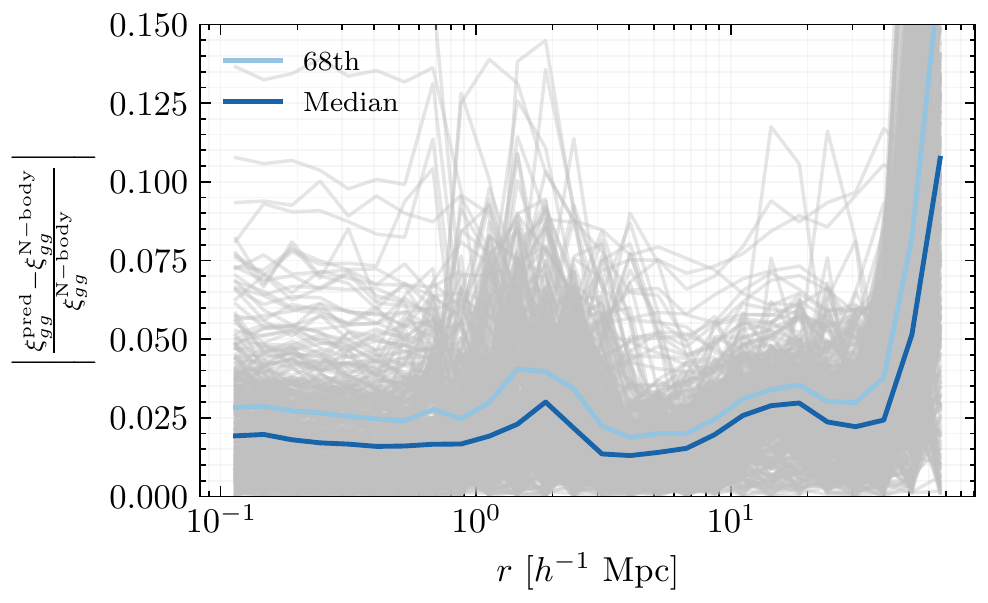}
    \caption{We show the absolute error of the emulator as a function of scale for each of the $400$ HOD mocks generated to test the accuracy of galaxy clustering predictions for different cosmologies, redshifts, and galaxy number densities. The light and dark blue lines show the 68th credible interval and the median of the absolute errors.} 
    \label{fig:galaxies_eval}
\end{figure}

Fig.~\ref{fig:galaxies_eval_z} shows the emulator predictions for $20$ HOD mocks at fixed redshift ($z=0.25$), each of the curves is generated from a different set of cosmological parameters in the test and validation sets. Comparing the mock HOD catalogues with the emulator predictions, we find that the median error of the emulator is below $3$ per cent on scales smaller than $50 \, \, h^{-1}\, \mathrm{Mpc}$, as shown in Fig.~\ref{fig:galaxies_eval}. Furthermore, the 68th percentile interval of the error increases only by $1$ per cent point with respect to the median. There is a small increase ($\approx 1$ per cent point) in the error in the transition from one-to-two-halo term that occurs between $1$ and $2 \, h^{-1}\, \mathrm{Mpc}$. On large scales, the variance of the measurements is large, making it difficult to accurately determine the error of the emulator.

Fig.~\ref{fig:error_trends_gals} shows the performance of the emulator as a function of the galaxy number density and redshift. In both cases, the emulator shows similar levels of performance and therefore does not show any bias.

\section{Solving the inverse problem: From correlations to cosmology}
\label{sec:inverse_problem}
Here, we show how the galaxy two-point correlation function emulator is able to recover the cosmological parameters from mock simulated galaxies, first using the same HOD prescription as the one implemented in our theoretical model within the $68\%$ credible interval for all parameters. 

It should be emphasised that we focus on the three-dimensional two-point correlation of galaxies in real space, which is not directly observable in galaxy surveys. What we observe is the redshift space two-point correlation function of galaxies, which will be the subject of the second paper in this series. However, it is important to show that the emulator is capable of recovering the parameters of interest for a mock dataset and to study the potential biases that might arise from adopting a too simplistic HOD model. We will also examine the scale dependence of the cosmological information content, which will, in turn, be important in determining the information content in redshift space.

We generated mock galaxy catalogues for LOWZ SDSS-like galaxies based on the fiducial Planck cosmology of the \textsc{Dark Quest} HR simulations, following \citet{2020PhRvD.101b3510K}. See Table~\ref{table:hod_params} for the characterisation of the mock sample.

\begin{table*}
\centering
\caption{The fiducial values and priors of the parameters for mock galaxy surveys that resemble the LOWZ galaxy sample.}
\label{table:hod_params}
\begin{tabular}{c c c c c c c c}
 \hline
 \hline
  & $\bar{z}$ & $\bar{n}_{\rmg}$ $\left[(h^{-1}\mathrm{Mpc})^{-3}\right]$ & $\log M_{\mathrm{min}}\, [h^{-1}M_{\odot}]$  & $\sigma_{\log{M}}$ & $\log M_1 \, [h^{-1}M_{\odot}]$ & $\kappa$ & $\alpha_{\mathrm{sat}}$ \\ 
  \hline
 \hline
  Fiducial & 0.251 & $2.174\times10^{-4}$ & 13.62 & 0.6915 & 14.42 & 0.51 & 0.9168 \\  
  \hline
  Min prior & - & - & 12 & 0.1 & 12 & 0.01 & 0.5 \\  
  \hline
  Max prior & - & - & 14.5 & 1 & 16 & 3 & 3 \\  
  \hline
\end{tabular}
\end{table*}

We use nested sampling, in particular the implementation of \textsc{pymultinest} \citep{2014A&A...564A.125B}, to obtain samples from the posterior distribution. The posterior is defined as 
\begin{equation}
    p(\theta|\mathcal{D}) \propto \mathcal{L}(\mathcal{D}|\theta) p(\theta),
\end{equation}
where $\theta$ are the parameters to be estimated, $p(\theta|\mathcal{D})$ is the posterior distribution of the parameters given the data, $\mathcal{L}(\mathcal{D}|\theta)$ describes the likelihood of the data given the parameters, and $p(\theta)$ is the prior distribution of the model parameters. 

We used a combination of the real space two-point correlation function and galaxy number density as our data vector and assumed that the likelihood follows a Gaussian distribution. Therefore, we compute the log-likelihood (up to a normalisation factor) as follows
\begin{equation}
\begin{split}
\label{eq:likelihood}
    \mathcal{L}(\mathcal{D}|\theta) & = -\frac{1}{2} \sum_{r_i, r_j} \left[\xi^s(r_i) - \xi^s(r_i|\theta) \right] \times C^{-1}(\xi^s(r_i), \xi^s(r_j)) \\ & \times \left[\xi^s(r_j) - \xi^s(r_j|\theta) \right] + \frac{(n_{\rmg}^s - n_{\rmg}^s(\theta))^2}{\sigma^2_{n_{\rmg} }},
\end{split}
\end{equation}
where $\xi^s(r_i)$ denotes the two-point correlation function of the data for sample $s$, and $\xi^s(r_i|\theta)$ is the prediction of the theoretical model where $\theta$ denotes the model parameters, i.e.  cosmological and HOD ($\mathcal{C} + \mathcal{G}$), $C$ is the data covariance matrix for a volume of $0.67$ $(h^{-1} \,\mathrm{Gpc})^3$, $n_{\rmg}^s$ is the galaxy number density estimated from the data, $n_g^s(\theta)$ the theoretical prediction, and $\sigma_{n_{\rm g}}$ the estimated error of the data that we fix to a nominal value of $5$ per cent. Note that the galaxy number density depends both on the HOD parameters and on cosmology, as seen in Eq.~\eqref{eq:n_g}. See Appendix~\ref{sec:appendix_covariance} for a description of how the covariance matrix is estimated from N-body simulations. 

Unless otherwise stated we will use the entire range of scales on which the emulator was trained, $0.1 \, h^{-1}\,\mathrm{Mpc} \leq r \leq 150 \, h^{-1}\,\mathrm{Mpc}$, to perform inference. Furthermore, although we vary the cosmological parameters $\mathcal{C} = \{ \Omega_\Lambda,\ln A_{\rm s}, \omega_{\rm c} \}$, we show constraints on the derived parameters most commonly used $\mathcal{C} = \{ \Omega_{\mathrm{m}}, \sigma_8, h \}$. The priors on the cosmological parameters are chosen to be uniform within the range of the sampled latin hyper-cube (Eq.~\ref{eqn:dq_param_range}); the priors on the HOD parameters are also chosen to be uniform with the ranges shown in Table~\ref{table:hod_params}.

\subsection{Fiducial constraints}
\begin{figure*}
    \centering
    \includegraphics[width=0.9\textwidth]{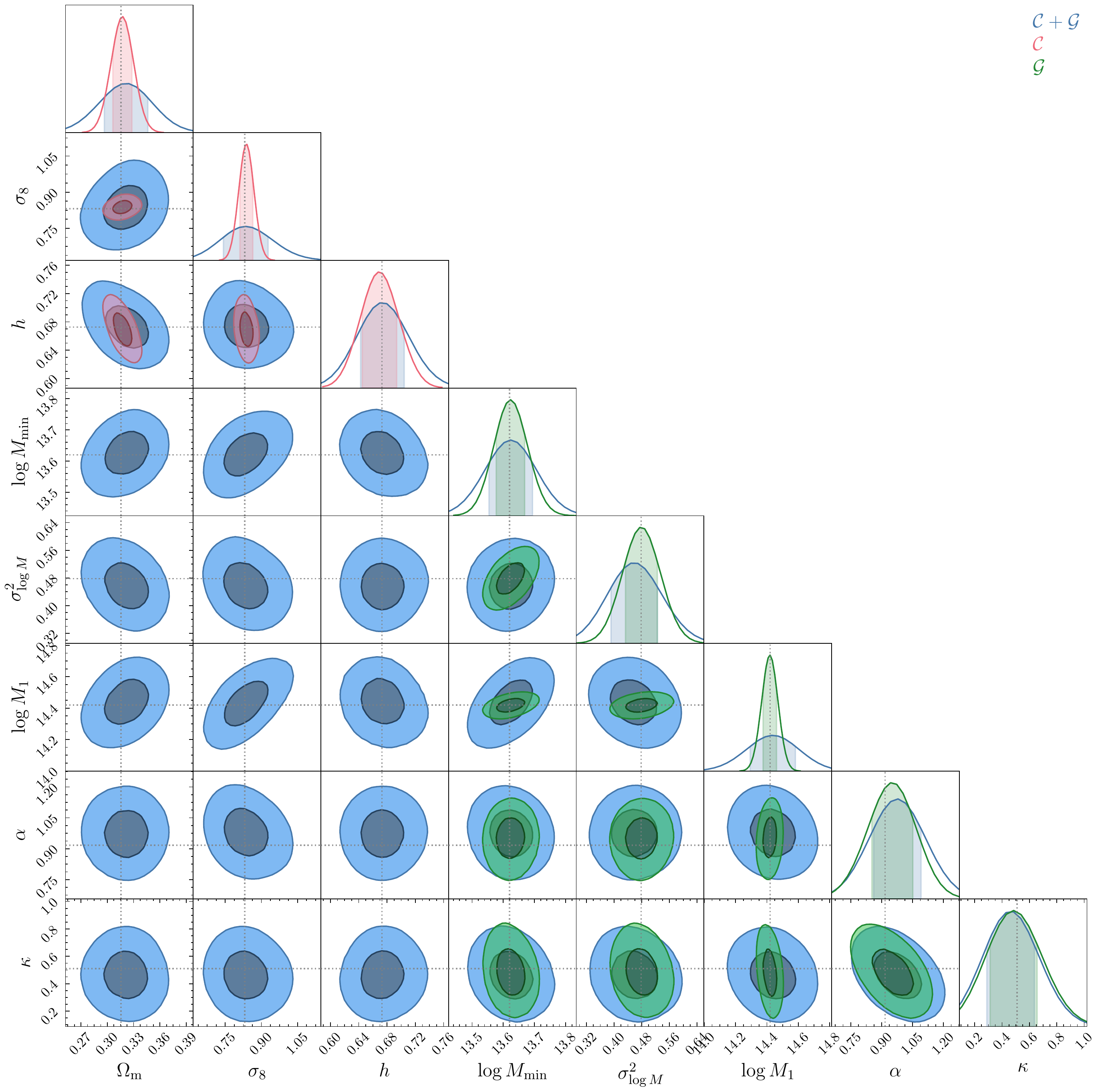}
    \caption{This plot shows that the emulator can recover the true cosmological and HOD parameters within the confidence intervals. We show the posteriors which result when varying both cosmology and HOD parameters ($\mathcal{C}$ and $\mathcal{G}$) (blue, labelled ``$\mathcal{C}+\mathcal{G}$") and the cosmological constraints found when the HOD parameters ($\mathcal{C}$) are set to their fiducial values (red, labelled ``$\mathcal{C}$"). The constraints on the HOD parameters ($\mathcal{G}$) obtained by fixing the cosmological parameters to their fiducial values are shown in green (labelled ``$\mathcal{G}$"). The true values that generated the simulated data are shown by the dotted gray lines.}
    \label{fig:fiducial_fit}
\end{figure*}

Here, we show that the emulator is capable of recovering the fiducial parameters of the mock catalogue within the $68\%$ confidence interval for all parameters. The resulting 2-D posterior distributions are shown in blue in Fig.~\ref{fig:fiducial_fit}. 

In the same figure, we also show the resulting constraints when the HOD parameters are fixed to their fiducial values (green) and the constraints on the HOD parameters when the cosmological parameters are fixed to their fiducial values (red). 

Although taking either of these two steps in a real analysis would underestimate the error on the estimated parameter values, and most likely bias them, this is a useful exercise to determine how much more one could learn by combining the two-point correlation function with other statistics that can constrain the HOD parameters more accurately. For example, \citet{Hahn_2021} demonstrated how using the bispectrum could help us to improve constraints on both the cosmological and HOD parameters, by breaking degeneracies between them. Other probes, such as galaxy-galaxy weak lensing \citep{More_2015} can also be used to infer the HOD parameters. Fig.~\ref{fig:fiducial_fit} shows that the constraints on $\Omega_{\rm m}$ and $\sigma_8$ could be significantly improved by breaking the degeneracies with the HOD parameters. 

On the other hand, it is mostly the mass scales $M_\mathrm{min}$ and $M_1$ that are better constrained by galaxy clustering when fixing the cosmological parameters. The remaining satellite parameters $\alpha$ and $\kappa$ do not improve significantly by fixing cosmology. This is probably due to the fact that LOWZ galaxies have a low fraction of satellites, compared with other galaxy selections, and therefore their galaxy two-point correlation function is not very sensitive to these two satellite occupation parameters.

Fig.~ \ref{fig:ngal_comparison_fit} shows the effect of removing the number density constraint from the likelihood. As previously found in \citet{2021arXiv210100113M}, the constraints on cosmological parameters are not strongly affected by the number density term. However, the HOD parameters are sensitive to this change, with the parameters that influence the number of centrals becoming much more poorly constrained when the number density is not used. 

\subsection{The complementary role of small scales}
\begin{figure*}
  \centering
  \begin{subfigure}{\textwidth}
    \centering
    \includegraphics[width=.98\textwidth]{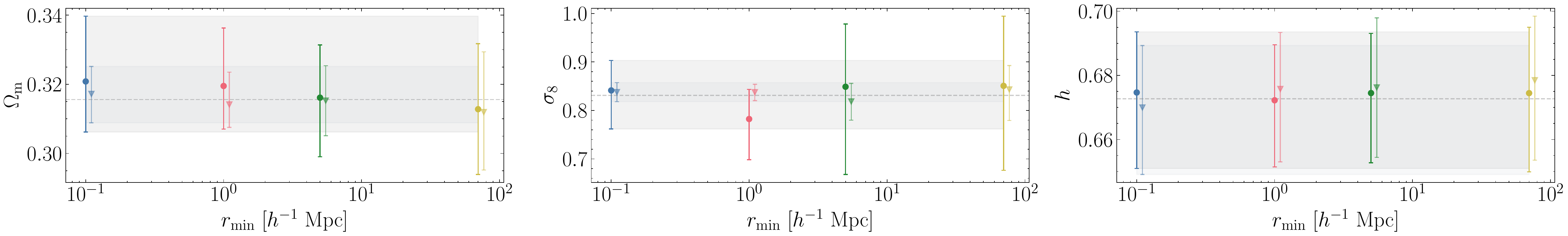}
  \end{subfigure}

  \begin{subfigure}{\textwidth}
    \centering
    \includegraphics[width=.98\textwidth]{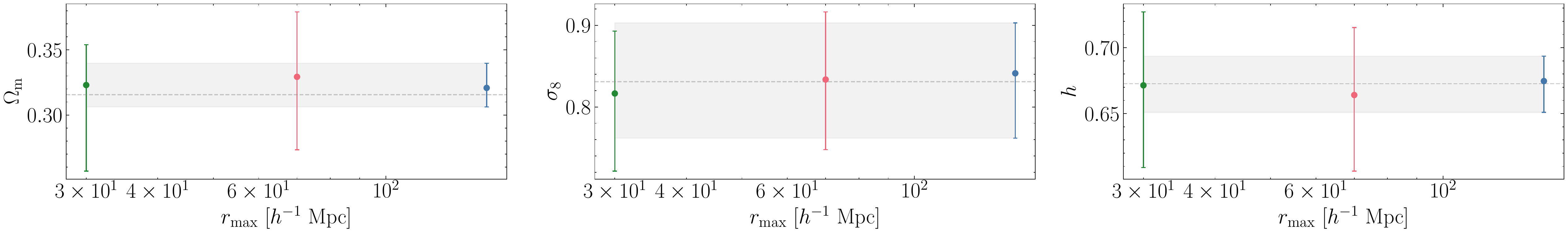}
  \end{subfigure}  
  \caption{We show the estimated maximum likelihood parameters, together with their estimated uncertainties, for varying minimum and maximum pair separation scales used in the analysis. In the top panel we show both the cosmological constraints obtained when marginalizing over the HOD parameters (circles) and when fixing the HOD parameters to their fiducial values (triangles). This shows that the constraints on the cosmological parameters improve as more non-linear scales are included for all parameters but $h$, whose constraints are dominated by the BAO information.}  
\label{fig:scale_dependence}
\end{figure*}

Here, we study how the constraints vary as a function of the minimum scale included in the likelihood evaluation. This is a test of the performance of our model and its accuracy on small scales, and serves to illustrate the usefulness of small scales in reducing the errors on the recovered parameters. We show the results of this test in Fig.~\ref{fig:scale_dependence}. 

The small-scale information mainly constrains the fluctuation amplitude, $\sigma_8$, as shown in the upper panel of Fig.~\ref{fig:scale_dependence}. From $r_\mathrm{min}=1 \, h^{-1}\,\mathrm{Mpc}$ to  $r_\mathrm{min}=5 \, h^{-1}\,\mathrm{Mpc}$, the errorbars on $\sigma_8$ increase by a factor of $~2$. 

In the same figure, we also show how the constraints on cosmological parameters would change if we fixed the HOD parameters. Interestingly, the $\Omega_{\rm m}$ constraints would also be improved by including small-scale information by about a factor of $2$ if there were no degeneracies with the HOD parameters. The constraints on $h$ are dominated by the BAO scale and therefore do not change noticeably when smaller scales are included or the HOD parameters are fixed.

In the bottom panel of Fig.~\ref{fig:scale_dependence}, we show the opposite effect, that of excluding large-scale information. The BAO scale has a very small effect on the recovered value of $\sigma_8$, whereas it dominates the constraints on the cosmological parameters $\Omega_{\rm m}$ and $h$, after marginalising over the HOD parameters. Note that most emulators \citep{Zhai_2019,2022arXiv220311963Y} focus on scales smaller than $30 \, h^{-1}\,\mathrm{Mpc}$, and therefore lose constraining power on $\Omega_m$ and $h$.

\label{sec:scale_dependence}

\subsection{The consequences of ignoring assembly bias}
\label{sec:ab}
We now test whether the halo-connection model used here is flexible enough to obtain unbiased cosmological constraints when modelling the clustering of a sample known to contain assembly bias. Although dark matter halo mass correlates strongly with galaxy clustering, we know that dark matter halos experience different assembly histories even at fixed halo mass, and can display different clustering. These different assembly histories influence secondary properties of halos, and this, in turn, might also affect the formation of galaxies, and hence result in different galactic contents for halos of the same mass.

These effects are known as \textit{halo} and \textit{galaxy assembly bias}. Note that although these two effects share the word bias, they refer to different effects
\begin{itemize}
    \item \textit{Halo assembly bias} refers to differences in the clustering of dark matter halos at a fixed halo mass. These differences depend on the choice of secondary halo properties, which usually correlate with the formation history of the halo, such as halo concentration or substructure fraction.
    \item \textit{Galaxy assembly bias} refers to differences in the number of galaxies within dark matter halos at a fixed halo mass, which in turn may depend on secondary halo properties.
\end{itemize}
Galaxy clustering is shaped by both of these effects. On one hand, halo assembly bias implies that, at fixed halo mass, grouping dark matter halos by a secondary property results in a different clustering signal. On the other hand, the way galaxies occupy dark matter halos might depend on properties other than mass. The combination of both effects determines how strongly galaxy clustering depends on secondary dark-matter halo properties, and therefore how important it is to model this dependency in order to obtain unbiased cosmological constraints. 

Here, we want to test how assembly bias affects our constraints when we include effects similar to those observed in hydrodynamical simulations \citep{2021MNRAS.508..698H} and semi-analytical models of galaxy formation \citep{Zehavi:2018, 2021MNRAS.502.3242X, Jimenez:2021} in our mock galaxy catalogues. In this way, we can assess whether the halo model is flexible enough to recover unbiased constraints from realistic galaxy mocks when including small-scale information. 

In particular, we implement the assembly bias model based on environment introduced in \citet{2021MNRAS.502.3242X}. The authors showed that the smoothed matter density can account for most of the assembly bias signal observed in a semi-analytic galaxy formation model. This is in agreement with other studies using hydrodynamical simulations \citep{2021MNRAS.508..698H}.

To create mock galaxy catalogues with an environment-based assembly bias signal, we first determine the local density around each halo. We compute the dark matter density field smoothed with a Gaussian filter over a scale of $2.5 \, h^{-1}\mathrm{Mpc}$, by first measuring the counts-in-cell dark matter particle density on a $512^3$ grid and then multiplying with a Gaussian kernel in Fourier space. The matter overdensity value at the position of each halo is found by interpolating over the 3D grid. Finally, we rank the overdensity values of the halos at fixed halo mass and normalise them to be between $0$ and $1$. Note that we have computed the ranks inside $50$ logarithmically spaced halo mass bins in the range $12 < \log_{10} \qty[M_{\mathrm{h}} / (h^{-1} M_{\odot})] < 16$. These ranks, $\delta^\mathrm{rank}_{2.5}$, are then normalised between 0 and 1 in each halo mass bin. 

Once we have determined the ranked environment density around each halo, we assign galaxies to dark matter halos through equations Eq.~\eqref{eq:occ_centrals} and Eq.~\eqref{eq:occ_satellites}, modifying the values of $ \log M_\mathrm{min}$ and $\log M_1$ with the rank of the halo's overdensity value
\begin{equation}
    \log_{10} M_\mathrm{min} (\delta^\mathrm{rank}_{2.5}) = \log_{10} M_\mathrm{min}^0 + B_\mathrm{cen} \times \left( \delta^\mathrm{rank}_{2.5} - 0.5 \right),
\end{equation}
\begin{equation}
    \log_{10} M_1 (\delta^\mathrm{rank}_{2.5}) = \log_{10} M_1^0 + B_\mathrm{sat} \times \left( \delta^\mathrm{rank}_{2.5} - 0.5 \right),
\end{equation}
where $B_\mathrm{cen}$ and $B_\mathrm{sat}$ are the central and satellite assembly bias parameters that control the strength of the effect. Since more galaxies will form in overdense regions, the values of $B_\mathrm{cen}$ and $B_\mathrm{sat}$ will be negative.

To explore the possible biases that ignoring assembly bias may introduce in the estimated cosmological parameters, we study two scenarios: i) a weak assembly bias effect with values $B_\mathrm{cen} = -0.1$ and $B_\mathrm{sat}=-0.2$, and ii) a strong one with values $B_\mathrm{cen} = -0.2$ and $B_\mathrm{sat}=-0.4$. The weak assembly bias parameters have been chosen to mimic the level of assembly bias signal found in \citet{2021MNRAS.502.3242X} for a sample with a galaxy number density of $n_\mathrm{gal}=0.01$ $\left(h^{-1}\mathrm{Mpc}\right)^{-3}$. In Fig.~\ref{fig:ab_extra}, we show that the weak scenario produces changes in the two-point correlation function of up to $10$ per cent compared with the case with no assmebly bias, while the strong case increases the clustering by up to $20$ per cent.

\begin{figure}
    \includegraphics[width=0.45\textwidth]{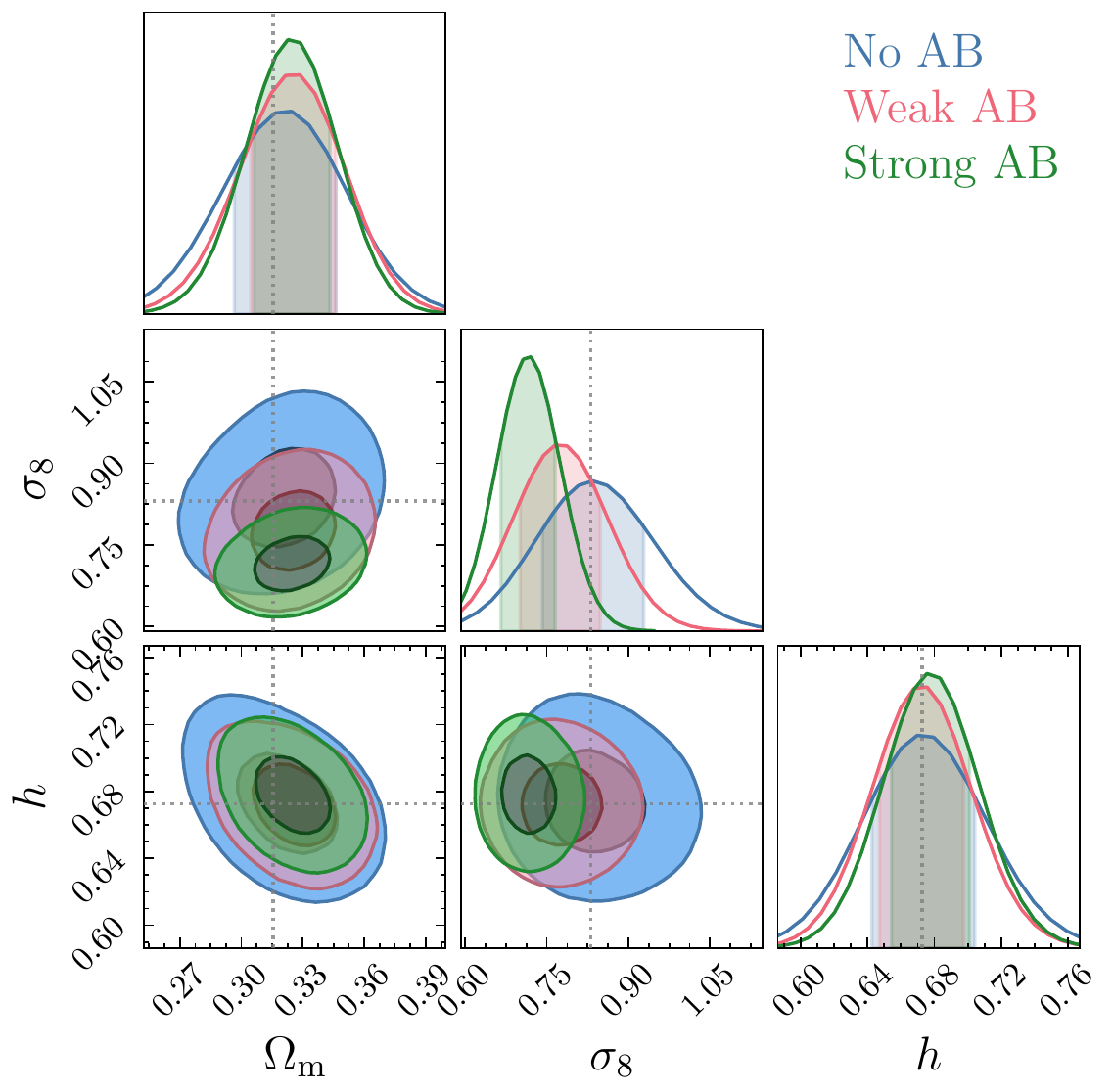}
    \caption{Constraints obtained when fitting mock catalogues that include the environment-based assembly bias model presented in \citet{2021MNRAS.502.3242X} with our halo model emulator, which ignores the effect of assembly bias. The cosmological parameters $\Omega_{\rm m}$ and $h$ can still be recovered within the estimated confidence intervals, since they are mainly constrained by the BAO peak, whereas $\sigma_8$ shows a small bias towards smaller values in both the weak and strong assembly bias scenarios.}
    \label{fig:ab_fit}
\end{figure}

Fig.~\ref{fig:ab_fit} shows the constraints obtained using our model (which ignores assembly bias) to fit the clustering measured from the mock galaxy samples described above, with weak and strong assembly bias. In both the weak and strong assembly bias scenarios, we can robustly recover the cosmological parameters $\Omega_{\rm m}$ and $h$ since they are mostly determined by the BAO scale. However, $\sigma_8$ is biased towards smaller values in both scenarios. In the strong assembly bias case, this shift is more than $1-\sigma$ away from its true value. However, we note that the strong assembly bias scenario is unrealistic for a LOWZ-like sample of galaxies \citep{2022arXiv220311963Y}. 

Fig.~\ref{fig:ab_full_fit} shows the full 2D posterior, including the HOD parameters that have shifted in the expected direction. Intuitively, the environmt assembly bias effect leads to more galaxies forming in overdense regions (thus, the assembly bias parameters are negative). The left hand side of Fig.~\ref{fig:ab_extra} shows that higher number densities in the assembly bias mocks correspond to a higher mean number of galaxies, that could be effectively reproduced by lowering $M_\mathrm{min}$. 

\begin{figure}
    \includegraphics[width=0.45\textwidth]{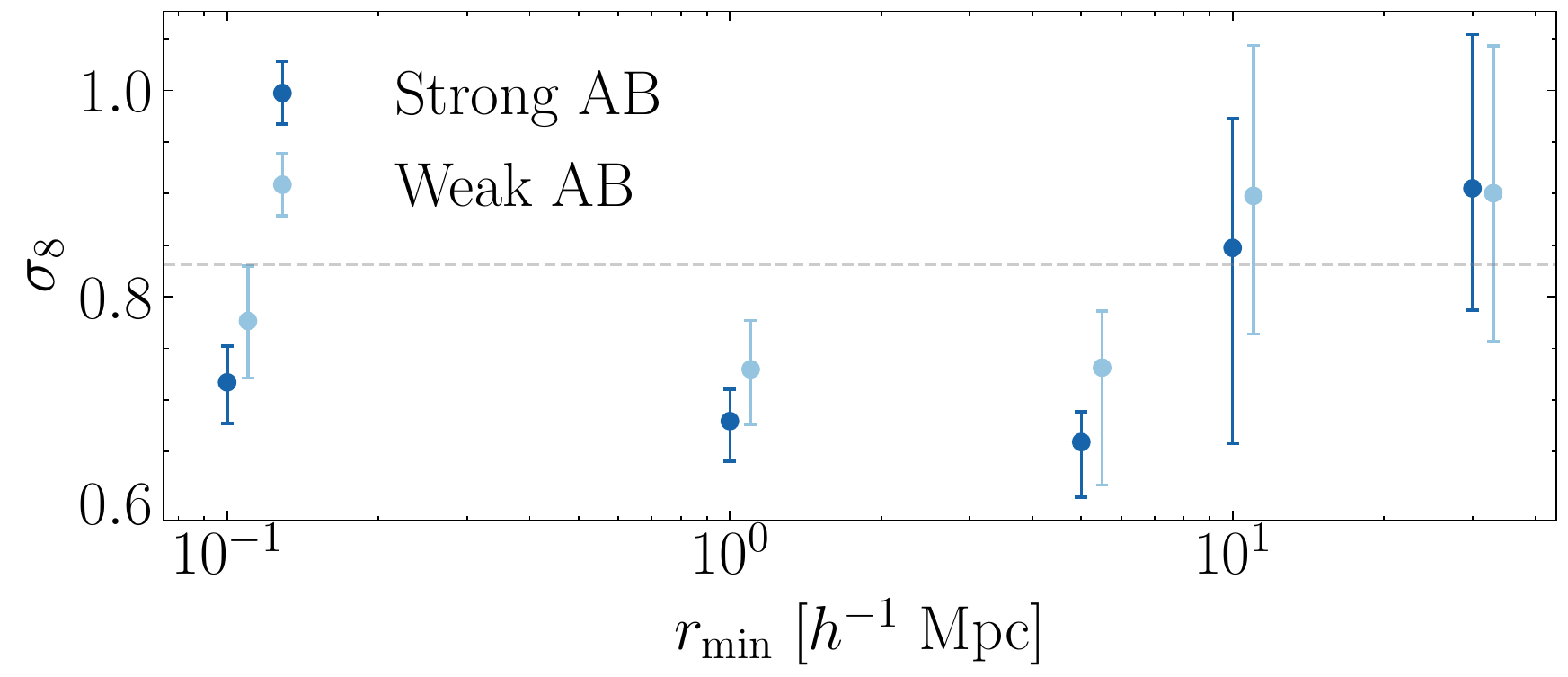}
    \caption{Inferred values of $\sigma_8$ and their estimated uncertainties as a function of the minimum scale, $r_\mathrm{min}$, used in the likelihood analysis. This plot shows the systematic introduced by assembly bias can only be removed by excluding the small scale information.}
    \label{fig:ab_scale}
\end{figure}

Fig.~\ref{fig:ab_scale} shows how the constraints on $\sigma_8$ change as we vary the minimum scale included in the determination of the likelihood. If we restrict the analysis to scales larger than $10$ $h^{-1} \,\mathrm{Mpc}$, the halo model recovers unbiased cosmological constraints by biasing the HOD parameters. However, on scales smaller than $10$ $h^{-1} \,\mathrm{Mpc}$, when the constraining power on $\sigma_8$ doubles, lowering the mass of halos that host a central cannot mimic the effects shown in Fig.\ref{fig:ab_extra}, and $\sigma_8$ needs to be lowered to describe the changes around the one to two halo term transition. 

We can monitor the evidence of the model to detect whether the halo-galaxy connection model has been mispecified. The evidence is defined as 
\begin{equation}
    P(\mathcal{D}) = \int \rm d \theta \, P(\mathcal{D}|\theta) P(\theta),
\end{equation}
and can be interpreted as the likelihood of the data given the model. The values of the evidence estimated by nested sampling are $20.87$ for mocks without assembly bias, $18.34$ for those with a weak assembly bias signal, and $16.37$ for those with a strong assembly bias effect. 

Given the importance of unbiased constraints on $\sigma_8$ to resolve the $\sigma_8-S_8$ tension, we will work on adding environment-based assembly bias to our emulator for its application to DESI Y1 data.

\subsection{Comparison with Lagrangian Perturbation Theory}
In this section, we compare the emulator constraints with those obtained by 1-loop Lagrangian perturbation theory \citep{2020JCAP...07..062C,2021JCAP...03..100C} using the publicly available code \textsc{velocileptors}\footnote{\url{https://github.com/sfschen/velocileptors}}. We fit the bias parameters $b_1$, $b_2$, and $b_s$, together with the cosmological parameters. We find that a LOWZ-like sample in real space cannot constrain the one-loop effective field theory counter-terms and therefore we set them to zero.

\begin{figure}
    \includegraphics[width=0.45\textwidth]{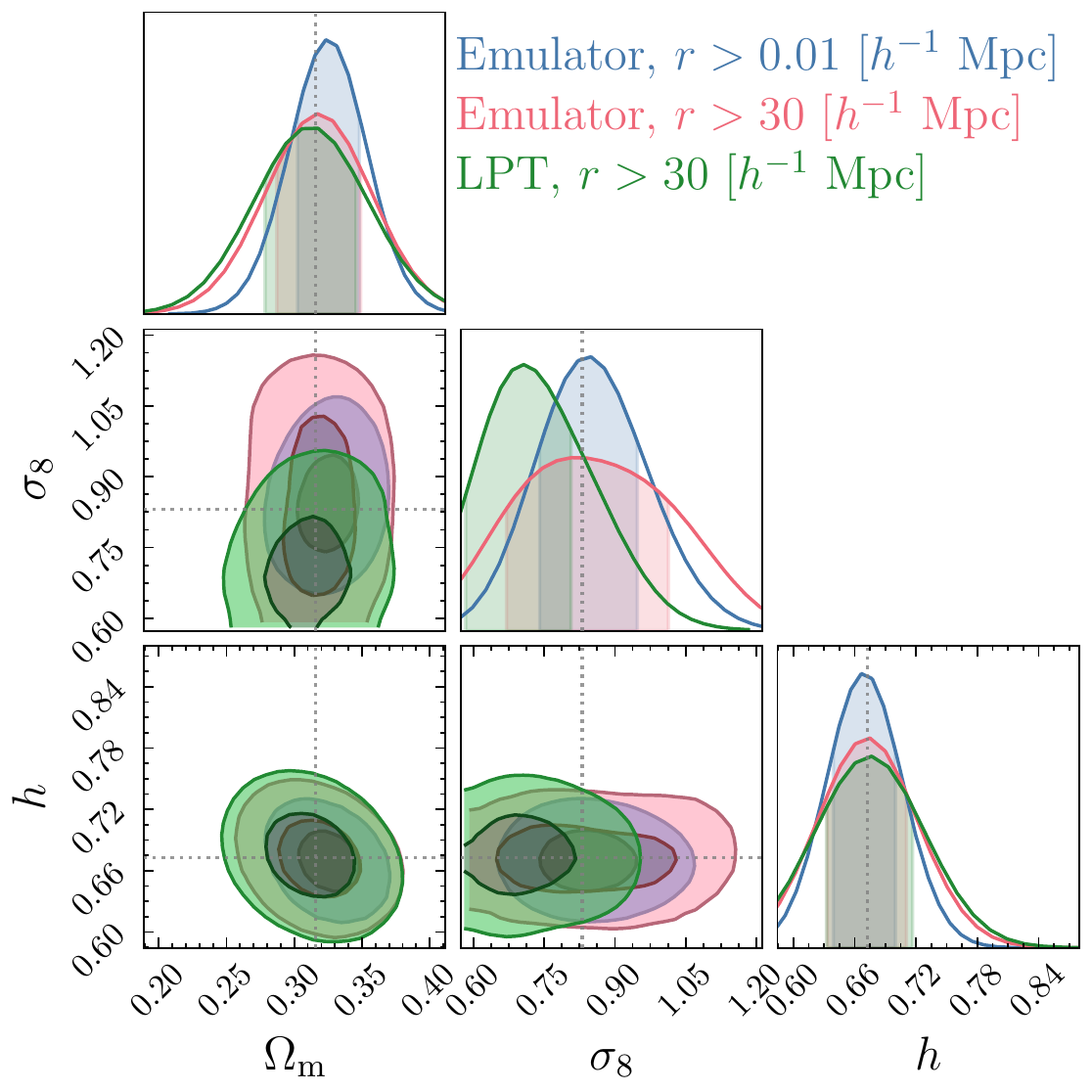}
    \caption{Comparison of the constraints obtained by the emulator based model to the 1 loop Perturbation Theory model presented in \citet{2020JCAP...07..062C,2021JCAP...03..100C}. }
    \label{fig:eft_comparison}
\end{figure}

In Fig.~\ref{fig:eft_comparison}, we show how the emulator can obtain constraints similar to LPT when analysed over the same scale range, even after marginalising the halo-galaxy connection parameters, which are in total $6$ free parameters (compared to only $3$ for LPT). Note that the LPT predictions are slightly biased in $\sigma_8$, this is due to the strong degeneracy between $b_1$ and $\sigma_8$ that is accentuated in real space. In such a situation, the 1-D marginalized posterior for $\sigma_8$ can depend strongly on the prior or the parameterisation of the nuisance parameters, potentially leading to a biased estimate \citep{2020PhRvD.102h3520S}. The biased estimate of $\sigma_8$ tends to be alleviated by including more information, e.g., redshift space distortions. As shown in Fig.~\ref{fig:eft_comparison}, including small scale information does allow the emulator to constrain the parameters more accurately.

\section{Discussion and Conclusion}
\label{sec:discussion}
We show that after marginalizing over uncertainties in the galaxy-halo connection parameters, an emulator of the real space correlation function based on the halo model can obtain tighter constraints on the cosmological parameters than Lagrangian Perturbation Theory (LPT) given that the latter cannot extract the additional information contained in small scale galaxy clustering. 

The treatment of galaxy bias in both approaches is very different. On the one hand, the bias treatment of LPT is based on expanding the galaxy number density perturbation, $\delta_{\rmg}(\boldsymbol{x})$, in terms of a series of local operators \citep{Desjacques_2018}, which are meant to capture the effect of the large-scale environment on the formation and evolution of galaxies. Each operator is associated with a free coefficient, called bias parameter, which depends on the selected population of galaxies and needs to be fitted to the data. The number and type of operators up to a given order in perturbation theory can be fully determined by symmetry considerations \citep{McDRoy0908,ChaSco1204,AssBauGre1408,Sen1511,MirSchZal1507,Desjacques_2018,EggScoSmi1906}, which guarantees that within its regime of validity the perturbative bias expansion can model any galaxy-matter connections (including assembly bias). On the other hand, the HOD approach implemented in this paper has the advantage that it can be extended further into the non-linear regime compared to the perturbative expansion, but is restricted by the assumption one makes about the halo properties that determine halo clustering and galaxy occupations. More work is needed to determine the robustness of both approaches against uncertainties in the model connecting halos to galaxies as well as their constraining power on cosmological parameters. In the future, we plan to compare the constraints obtained with both models using large hydrodynamic simulations or semi-analytic models of galaxy formation.

Regarding the emulation approach, we have combined an emulator trained in halo properties with an analytical prescription of how galaxies populate halos, as already done by \citet{Nishimichi:2018etk}. Most other emulators, however, are trained on HOD catalogues built on N-body simulations \citep{Zhai_2019,2022arXiv220311963Y}. Our approach has advantages and disadvantages. In particular, the halo model allows us to reduce emulator errors through an analytical galaxy-halo connection, which also simplifies the task for the emulator that only needs to learn the dependency of halo clustering on cosmological parameters. Moreover, the analytical model allows us to compute different observables, such as the galaxy-cluster cross-correlation function or a multitracer two-point correlation function. Obtaining cosmological information from small scales through these observables will be the subject of future work. It also allows us to combine emulators trained on simulations with different resolutions to reduce cosmic variance on large scales and perform an analysis using the full-shape of the correlation function.

Regarding the disadvantages of our approach, extending the halo model approach to arbitrary statistics could potentially be difficult. The emulation of statistics such as the bispectrum, would be simplified if one were to follow the procedure outlined in \citet{Zhai_2019,2022arXiv220311963Y}. Moreover, more work needs to be done in order to go beyond the vanilla HOD model used in this work to introduce effects such as the environment-based assembly bias shown in Section ~\ref{sec:ab}. In the future, we plan to introduce a correction based on binning the halo two-point correlation function in terms of halo environment.

We have shown that including environment-based assembly bias in the model is important to avoid biased constraints on $\sigma_8$. This is especially relevant given the $f\sigma_8$ tension. Previously, \citet{Kobayashi:2021oud} and \citet{2021arXiv210100113M} had performed tests similar to the one presented in Sec.~\ref{sec:ab} to emulators based also on the halo model. \citet{Kobayashi:2021oud} studied the effect that ignoring concentration-based assembly bias would have on the cosmological parameters inferred when emulating the redshift space power spectrum through the halo model. They found that although the mock galaxies show $10-20$ per cent higher amplitudes than the mocks without assembly bias, they can still recover unbiased cosmological constraints through a change in the HOD parameters. In contrast, \citet{2021arXiv210100113M} found that the same effects of assembly bias would introduce biases in $\Omega_{\rm m}$ and $\sigma_8$ when the data vector is a combination of the projected two-point correlation function of galaxies and galaxy-galaxy lensing. In this case, the fact that one can use galaxy-galaxy lensing to accurately determine the scaling of halo bias with halo mass restricts the flexibility of the HOD model, which is not able to adapt the parameters in such a way that unbiased constraints can be recovered. 

We have here explored an assembly bias model inspired by semi-analytic methods of galaxy formation and hydrodynamical simulations. In fact, these studies find that the magnitude of concentration-based assembly bias is small. Ignoring environment-based assembly bias in the theory model, we find that the halo model is not flexible enough to obtain unbiased cosmological constraints already when the effect of assembly bias only impacts clustering by about $10\%$. Moreover, we find that including the BAO scale allows us to obtain robust constraints on $\Omega_{\rm m}$.


To summarise, we have
\begin{itemize}
    \item Presented a neural network which models the full-shape galaxy clustering in real space based on the halo model, which is more accurate and faster than previously published Gaussian process emulators \citet{Nishimichi:2018etk}, when trained on the same dataset. The method presented here can produce a galaxy correlation function in less than $300$ ms on a single core.
    \item Shown that small scale galaxy clustering ($r < 5$ $h^{-1} \,\mathrm{Mpc}$) in real space improves the constraints on $\sigma_8$ by a factor of $~2$, whereas marginalising over the HOD parameters erases the information contained on small scales for $\Omega_{\mathrm{m}}$.
    \item Shown that a halo model that ignores effects of environment-based assembly bias similar to those observed in hydrodynamic simulations and semianalytic models of galaxy formation could introduce bias in the inferred $\sigma_8$, while the BAO peak ensures that we can recover $\Omega_{\rm m}$ and $h$ robustly.
    \item Found that the above-mentioned bias in the value of inferred $\sigma_8$ disappears when analysing scales larger than $10$ $h^{-1}\,\mathrm{Mpc}$.
\end{itemize}
In the second paper of this series, we will present analogous neural network emulators of the pairwise velocity moments that will be used to i) perform the real to redshift space mapping to predict the cosmological dependence of redshift-space galaxy clustering, and ii) constrain observations of the peculiar velocity field. 

In the future, we also plan to use the neural network emulators on DESI Y1 data to constrain the cosmological parameters. This requires that the models be trained on simulations with lower particle mass so that they can reach the high galaxy number densities that DESI will measure. For this, a new simulation campaign, Dark Quest II., is currently ongoing to cover a wider mass range (down to a few $10^{11}\,h^{-1}M_\odot$) in an extended cosmological model space including massive neutrinos, time-varying dark energy equation-of-state parameter and spatial curvature using a newly developed fast $N$-body code (Nishimichi \textit{et al.} in prep.). 

\section*{Acknowledgements}


CC would like to acknowledge support from the Science Technology Facilities Council through a Centre for Doctoral Training in Data Intensive Science studentship (ST/P006744/1). AE is supported at the AIfA by an Argelander Fellowship. SB is supported by the UK Research and Innovation (UKRI) Future Leaders Fellowship [grant number MR/V023381/1]. This work was also supported by STFC grant ST/T000244/1. This work used the DiRAC@Durham facility managed by the Institute for Computational Cosmology on behalf of the STFC DiRAC HPC Facility (www.dirac.ac.uk). The equipment was funded by BEIS capital funding via STFC capital grants ST/K00042X/1, ST/P002293/1,
ST/R002371/1 and ST/S002502/1, Durham University and STFC operations grant ST/R000832/1. DiRAC is part of the National eInfrastructure. This work was also  supported in part by MEXT/JSPS KAKENHI Grant Number JP19H00677, JP20H05861, JP21H01081 and JP22K03634. We also acknowledge financial support from Japan Science and Technology Agency (JST) AIP Acceleration Research Grant Number JP20317829.

\section*{Data Availability}

The data shown in this article will be shared on reasonable request
to the corresponding author.



\bibliographystyle{mnras}
\bibliography{example} 




\appendix
\section{Evaluation of the emulators as a function of redshift and number density}
\begin{figure*}
    \centering
    \includegraphics{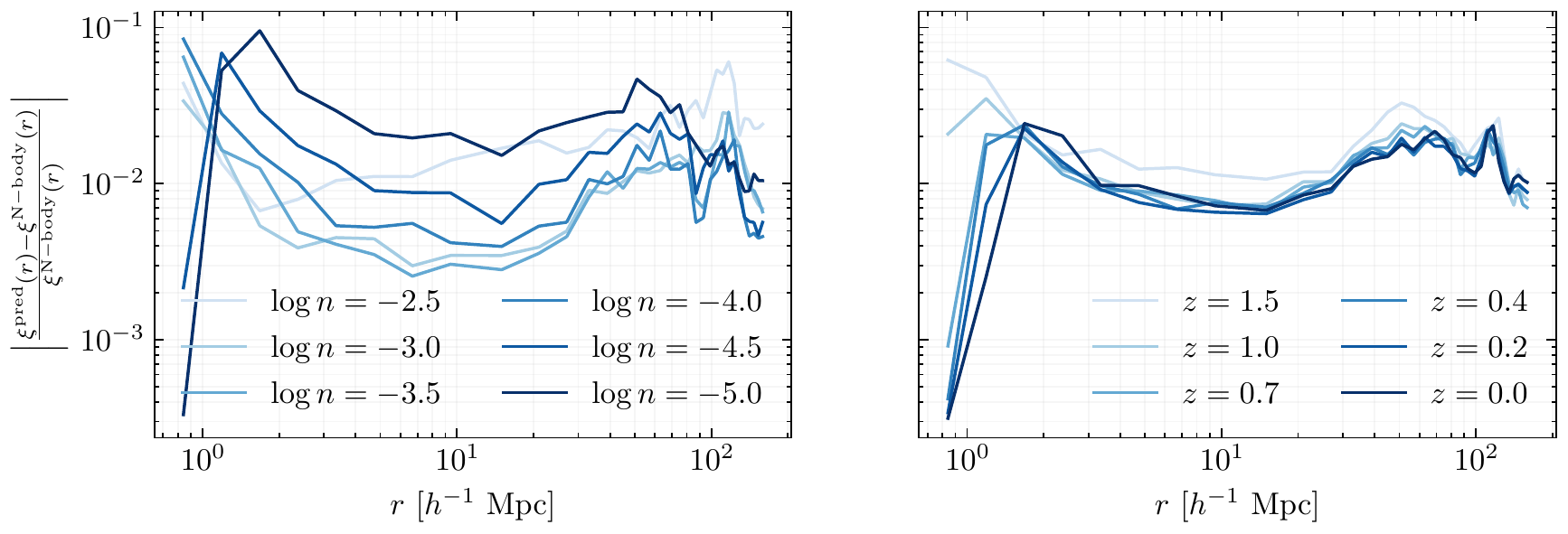}
    \caption{Median absolute errors of the halo two-point correlation function as a function of number density (left), averaged over redshift and test set cosmologies, and as a function of redshift (right), averaged over number density and test set cosmologies.}
    \label{fig:error_trends_halos}
\end{figure*}

\begin{figure*}
    \centering
    \includegraphics{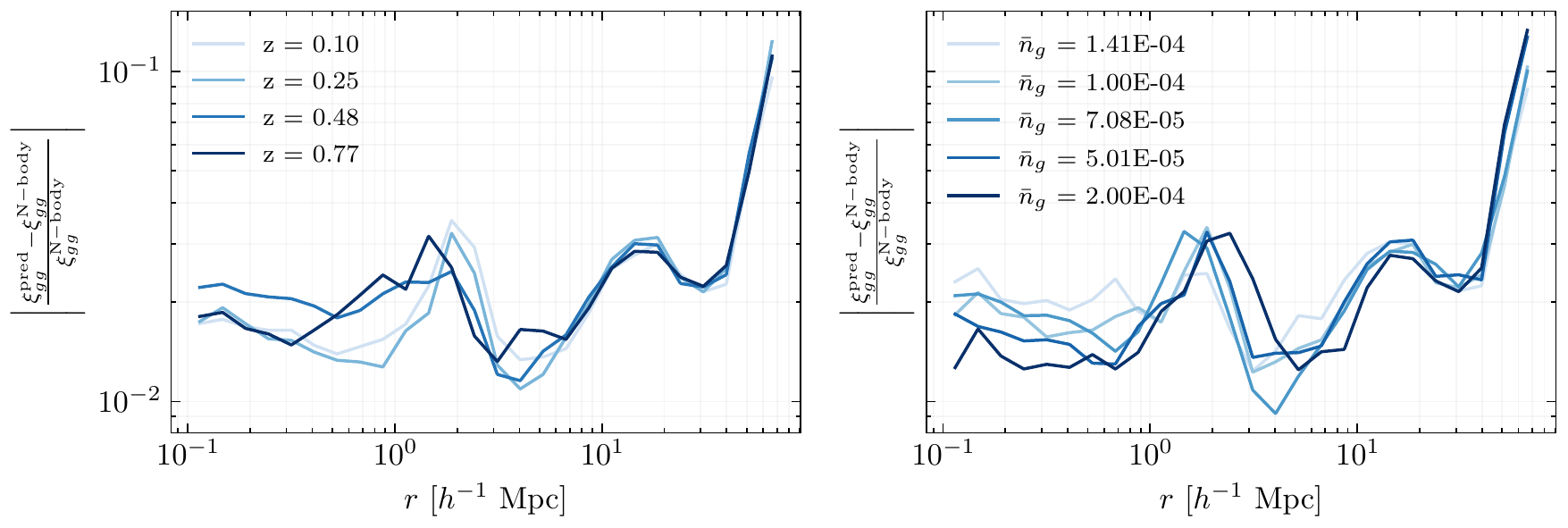}
    \caption{Median absolute errors of the galaxy two-point correlation function as a function of number redshift (left), averaged over galaxy number density and test set cosmologies, and as a function of galaxy number density (right), averaged over number redshift and test set cosmologies. In both cases the emulator accuracy does not show noticeable biases.
    }
    \label{fig:error_trends_gals}
\end{figure*}

In this appendix we show detailled evaluations of the halo auto-correlation emulator (Fig.~\ref{fig:error_trends_halos}) and the galaxy auto-correlation emulator (Fig.~\ref{fig:error_trends_gals}).

For halo auto-correlations, we find that the emulator accuracy decreaes for lower number densities, which are more affected by shot noise, whereas it decreases for high redshifts (~$z=1.5$). 

For galaxy auto-correlations we do not find any substantial biases for neither redshift or galaxy number density.

\section{Estimating the covariance matrix}
\label{sec:appendix_covariance}
In Section~\ref{sec:inverse_problem}, we used an estimate of the covariance matrix to obtain the posterior of cosmological parameters given a mock data vector. The covariance matrix was estimated from a set of $1600$ N-body simulations part of the AbacusSummit suite \citep{10.1093/mnras/stab2484}. These are high resolution small boxsize simulations ($L_\mathrm{box} = 500$ $h^{-1} \,\mathrm{Mpc}$). 

Given the small boxsize of the simulations, we re-scale the covariance by a factor of $0.5^3/0.67$ to estimate the expected errors for a LOWZ-like sample, whose effective volume is $0.67$ $(h^{-1} \,\mathrm{Gpc})^3$. We also correct the covariance estimated from the mocks with Eq. 56 in \citet{Percival_2021}.

\section{The effect of constraining galaxy number density in the likelihood analysis}
In this appendix, we show the effect of removing the galaxy number density term in Eq.~\ref{eq:likelihood}. 

\begin{figure*}
    \centering
    \includegraphics[width=0.8\textwidth]{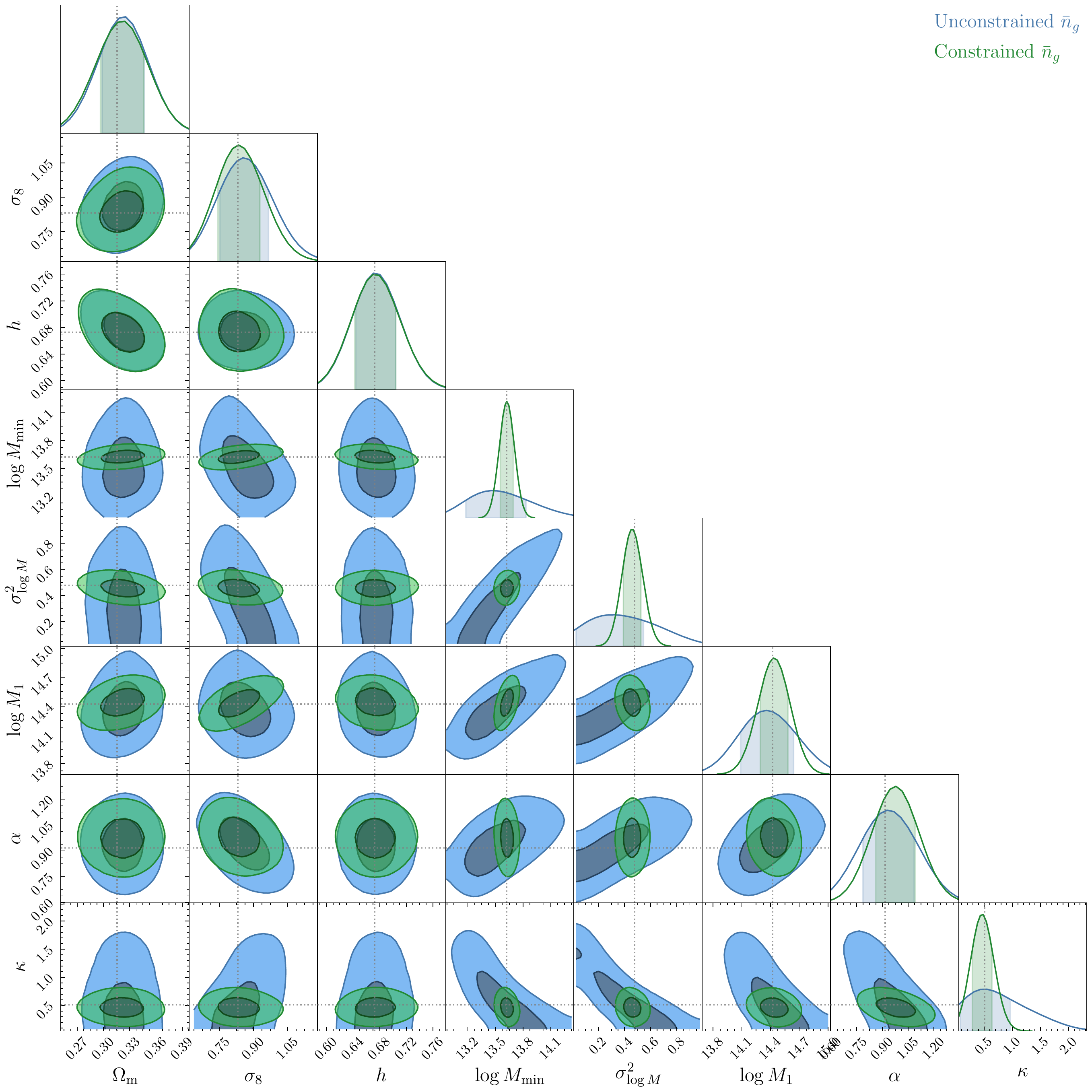}
    \caption{Comparison of constraints on cosmological and HOD parameters when the galaxy number density is included in the likelihood (Constrained $\bar{n}_g$) and when it isn't (Unconstrained $\bar{n}_g$. Including number density constraints only helps determine the HOD parameters with a higher accuracy.}
    \label{fig:ngal_comparison_fit}
\end{figure*}

Fig.~\ref{fig:ngal_comparison_fit} shows that the number density constrain does not change the constraints on cosmological parameters noticeably, whereas it mainly improves those of the HOD parameters. In particular, it breaks the degeneracy between the central occupation parameters, $\log M_\mathrm{min}$ and $\sigma_{\log{M}}$.

\section{Assembly bias mocks details}
Here, we describe here the occupation variations of the environment-based assembly bias mocks used in Section~\ref{sec:ab}. 

\begin{figure*}
     \centering
     \begin{subfigure}[b]{0.47\textwidth}
         \centering
         \includegraphics[width=\textwidth]{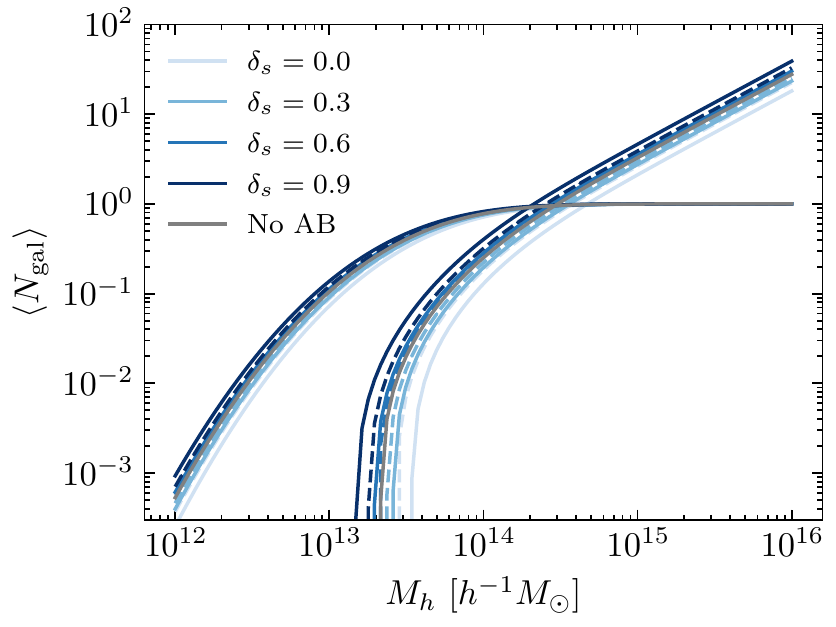}
     \end{subfigure}
     \hfill
     \begin{subfigure}[b]{0.47\textwidth}
         \centering
         \includegraphics[width=\textwidth]{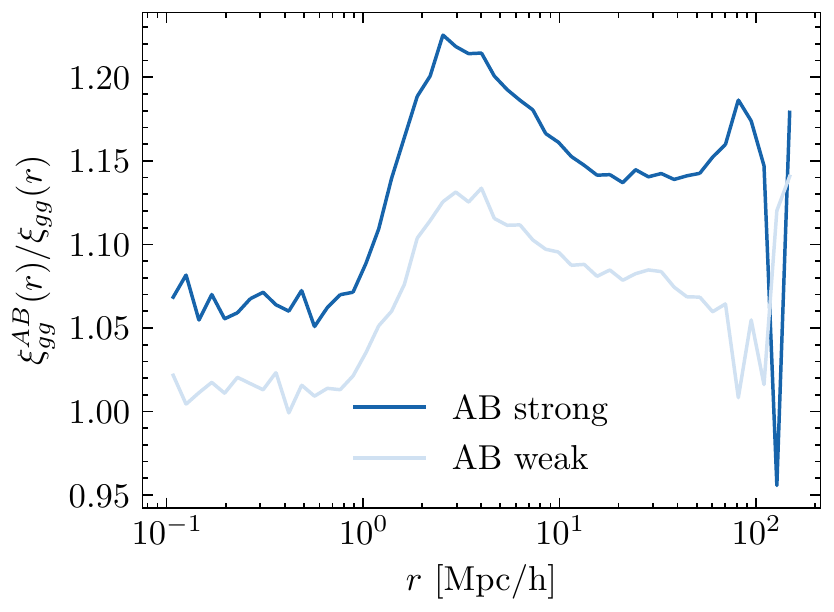}
     \end{subfigure}
     \hfill
    \caption{Details of the assembly bias mocks. On the left, we show the mean number of central and satellite galaxies as a function of halos mass and halo environment for both the strong assembly bias model (solid lines) and the weak assembly bias model (dashed lines). On the right, we show the ratios of the galaxy two-point correlation functions for assembly bias models, and their non-assembly bias counterpart.}
    \label{fig:ab_extra}
\end{figure*}

Fig.~\ref{fig:ab_extra} shows how the mean number of centrals and satellites change as a function of halo mass and halo environment for both the strong and weak assembly bias mocks. At fixed halo mass, halos residing in denser environments will have a higher mean number of galaxies (both centrals and satellites) than those occupying underdense regions. 

On the right hand side of Fig.~\ref{fig:ab_extra} we also show the ratio of the galaxy two-point correlation function with a strong and weak assembly bias signal to that of the no assembly bias case. The deviations can be as large as $10\%$ for the weak case, and $20\%$ for the strong one.

\begin{figure*}
    \centering
    \includegraphics[width=0.9\textwidth]{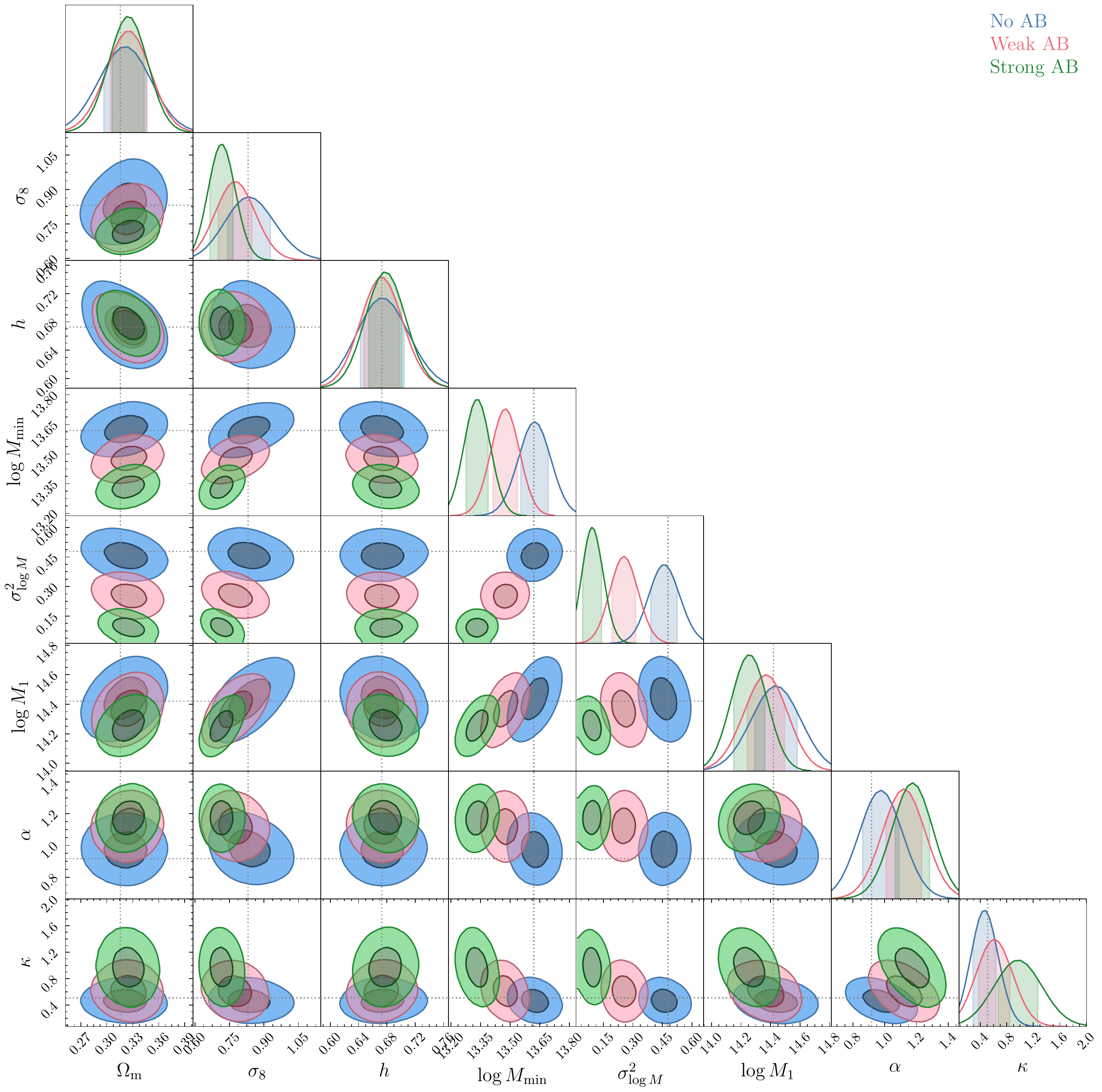}
    \caption{Full 2D posteriors obtained for data with i) no assembly bias effect, ii) a weak assembly bias signal, and iii) a strong assembly bias signal.}
    \label{fig:ab_full_fit}
\end{figure*}
\bsp	
\label{lastpage}
\end{document}